\documentclass[preprint,12pt]{elsarticle}

\usepackage{amssymb}

\usepackage{xr}
\usepackage{moreverb}
\usepackage{graphicx}
\usepackage{subcaption}
\usepackage{amsmath}

\usepackage[
colorlinks,bookmarksopen,bookmarksnumbered,citecolor=red,urlcolor=red]{hyperref}

\newcommand\BibTeX{{\rmfamily B\kern-.05em \textsc{i\kern-.025em b}\kern-.08em
T\kern-.1667em\lower.7ex\hbox{E}\kern-.125emX}}

\usepackage[font={bf, scriptsize}]{subcaption}
\usepackage{xspace}
\journal{Elsevier}

\begin{document}

\begin{frontmatter}



\title{Cross-validatory Z-Residual for Diagnosing Shared Frailty Models}


\author[inst1,inst2]{Tingxuan Wu}
\author[inst3,inst2]{Cindy Feng}
\author[inst1]{Longhai Li\corref{cor1}}

\affiliation[inst1]{organization={Department of Mathematics and Statistics, University of Saskatchewan},
            addressline={106 Wiggins Rd}, 
            city={Saskatoon},
            postcode={S7N5E6}, 
            state={SK},
            country={Canada}}

\affiliation[inst2]{organization={School of Public Health, University of Saskatchewan},
            addressline={104 Clinic Place}, 
            city={Saskatoon},
            postcode={S7N5E5}, 
            state={SK},
            country={Canada}}

\affiliation[inst3]{organization={Department of Community Health and Epidemiology, Faculty of Medicine, Dalhousie University},
            addressline={5790 University Ave}, 
            city={Halifax},
            postcode={B3H 1V7}, 
            state={NS},
            country={Canada}}
            
\cortext[cor1]{Corresponding author:}
\ead{longhai@math.usask.ca}

\begin{abstract}
Residual diagnostic methods play a critical role in assessing model assumptions and detecting outliers in statistical modelling. In the context of survival models with censored observations, Li et al. (2021) introduced the Z-residual, which follows an approximately normal distribution under the true model. This property makes it possible to use Z-residuals for diagnosing survival models in a  way similar to how Pearson residuals are used in normal regression.
However, computing residuals based on the full dataset can result in a conservative bias that reduces the power of detecting model mis-specification, as the same dataset is used for both model fitting and validation. Although cross-validation is a potential solution to this problem, it has not been commonly used in residual diagnostics due to computational challenges.
In this paper, we propose a cross-validation approach for computing Z-residuals in the context of shared frailty models. Specifically, we develop a general function that calculates cross-validatory Z-residuals using the output from the \texttt{coxph} function in the \texttt{survival} package in R.
Our simulation studies demonstrate that, for goodness-of-fit tests and outlier detection, cross-validatory Z-residuals are significantly more powerful and more discriminative than Z-residuals without cross-validation. We also compare the performance of Z-residuals with and without cross-validation in identifying outliers in a real application that models the recurrence time of kidney infection patients. Our findings suggest that cross-validatory Z-residuals can identify outliers that are missed by Z-residuals without cross-validation.
\end{abstract}




\end{frontmatter}

\footnotetext[1]{\textbf{List of Abbreviations}: AUC, area under the ROC curve; CHF, cumulative hazard function; CV, cross-validation; CS, Cox-Snell; GOF, goodness-of-fit; KM, Kaplan-Meier; LOOCV, leave-one-out cross-validation; QQ, quantile-quantile; SW, Shapiro-Wilk.}

\section{Introduction}

Residual diagnosis is a critical step in statistical modelling for checking the validity of model assumptions. Several residual diagnostic tools have been commonly used for checking the survival models \cite{collett_modelling_2015}, including Cox-Snell (CS) \cite{CoxD.R.1968AGDo}, martingale  \cite{therneau_martingale-based_1990}, deviance \cite{therneau_modeling_2013, mccullagh_generalized_1989}, Schoenfeld \cite{collett_modelling_2015, SCHOENFELDDAVID1982Prft} and scaled Schoenfeld \cite{ grambsch_proportional_1994-1} residuals. For example, the plot based on Cox-Snell residuals can be used as a graphical assessment tool for checking the overall goodness-of-fit (GOF) of a fitted model. The functional form of covariate is often examined using the plots of the martingale and deviance residuals against the covariates. The Schoenfeld and scaled Schoenfeld residuals are often used in testing the assumption of proportional hazards in the Cox proportional hazard model. Other residuals have also been proposed for diagnosing survival models \cite{ residual_mixturecuremodel_2017, residual_Tree-Structured_survival, residual_ph_interval_censored, deviance_normal_scores, lin_checking_1993, law_residual_2017,shepherd_probability-scale_2016-1, hillis_residual_1995}. However, there is a lack of residuals with a characterized reference distribution for censored regression. Li, et al.\cite{LiLonghai2021Mdfc} recently proposed the Z-residual diagnosis tool for diagnosing survival models with censored observations. The Z-residual is approximately normally distributed under the true model and has greater statistical power and is more informative than some traditional residual diagnostic tools for diagnosing model misspecifications, such as incorrect choice of distribution family and/or functional form of covariates of survival models.

The residuals considered in practical survival analysis are typically calculated based on the full dataset. When the same dataset is used to estimate the model parameters and calculate residuals for checking the fitted model, the power of detecting model misspecification may be reduced (bias) due to the double use of the dataset. The bias of the double use of the dataset has received much attention in the context of checking and comparing Bayesian models; see \cite{MarshallE.C.2003Acpc, MarshallE.C.2007IoiB,piironen_comparison_2017,vehtari_pareto_2015,vehtari_practical_2017,smith_prediction_2022,gelman_understanding_2014, LiLonghai2015Acpe,li_estimating_2017} and the references therein. For example,  Li, et al. \cite{LiLonghai2015Acpe} introduced integrated importance sampling methods for approximating leave-one-out cross-validatory predictive evaluations for models with unit-specific and possibly correlated latent variables. Cross-validatory predictive p-values can also be used to identify outliers by examining the tail probability of the predictive distribution \cite{MarshallE.C.2003Acpc, MarshallE.C.2007IoiB}. However, cross-validation is not commonly used in practical residual diagnosis in the frequentist paradigm. This is probably due to the computational challenges in cross-validation and the lack of awareness of the severity of the bias caused by the double use of the dataset. 

In this paper, we consider developing cross-validation methods for calculating the Z-residual for diagnosing survival models and comparing them to the Z-residual without cross-validation. We will focus on investigating the performance of cross-validatory Z-residuals in diagnosing shared frailty models.  A shared frailty model is a survival model by incorporating random effects (frailties) to account for unobserved heterogeneity \cite{vaupel_impact_1979}, where the frailties are shared among individuals within a cluster or group \cite{duchateau_frailty_2008, KaragrigoriouAlex2011FMiS, HanagalDD2015Msdu}. We develop a general R function for calculating the cross-validatory Z-residuals based on the outputs from fitting a survival model using the \texttt{coxph} function in \texttt{survival} package. We build an R function for splitting data into K-fold to ensure adequate representations of groups and other covariates in each fold. In our study design, the Z-residuals are calculated using three methods: the full dataset (No-CV), 10-fold cross-validation (10-fold) and leave-one-out cross-validation (LOOCV). We conduct simulation studies to investigate the performances of the three types of Z-residuals in detecting nonlinear covariate effects and identifying outliers through graphical visualization and SW tests. Our simulation results show that the SW tests based on 10-fold Z-residual and LOOCV Z-residual are significantly more powerful and more discriminative for detecting non-linear covariate effects. Moreover, cross-validatory Z-residuals are more powerful and more discriminative for identifying outliers than No-CV Z-residuals. In spite of these improved performances,  our simulation studies also show that the cross-validation results in elevated type-I error rates when cross-validatory Z-residuals are used in conducting GOF tests. Future research can be conducted to remedy this problem for cross-validatory Z-residuals.  We also compared the performance of the No-CV Z-residual and LOOCV Z-residual in identifying outliers for a kidney infection dataset \cite{MCGILCHRISTC.A1991RwFi}. The results show that the methods with LOOCV Z-residuals can identify some outliers that are missed by the No-CV method. 

The rest of this paper is organized as follows. Section \ref{sec:frailtymodel} gives a brief review of shared frailty models. In Section \ref{sec:cvz} we present the definition of the cross-validatory Z residual with a discussion of the algorithm for computing cross-validatory Z residuals. Section \ref{sec:simulation} presents the results of simulation studies for investigating the performances of 10-fold and LOOCV Z residuals. In section \ref{sec:realdata}, we present the results of applying the LOOCV Z residual to identify outliers for a kidney infection dataset. The article is concluded in Section \ref{sec:conc}.

\section{Shared gamma frailty models}\label{sec:frailtymodel}
A shared frailty model is a frailty model where the frailties are common or shared among individuals within groups. The formulation of a frailty model for clustered failure survival data is defined as follows. Suppose there are $g$ groups of individuals with $n_i$ individuals in the $i$th group, $i$ = 1, 2, \ldots , $g$. If the number of subjects $n_i$ is 1 for all groups, then the univariate frailty model is obtained \cite{KaragrigoriouAlex2011FMiS}. Otherwise, the model is called the shared frailty model \cite{henderson_analysis_2001, duchateau_frailty_2008, hougaard_frailty_1995}  because all subjects in the same cluster share the same frailty value $z_i$. Suppose $t_{ij}$ is the true failure time for the $j$th individual of the $i$th group, which we assume to be a continuous random variable in this article, where $j$ = 1, 2, . . . , $n_i$. Let $t_{ij}^{*}$ denote the realization of $t_{ij}$. In many practical problems, we may not be able to observe $t_{ij}^{*}$ exactly, but we can observe that $t_{ij}$ is greater than a value $c_{ij}$, where $c_{ij}$ be the corresponding censoring time. The observed failure times are denoted by the pair $(y_{ij}, \delta_{ij})$, where $y_{ij}=\min(t_{ij}, c_{ij}), \delta_{ij}=I (t_{ij} < c_{ij})$. The observed data can be written as $y=(y_{11},\ldots,y_{gn_g})$ and $\delta=(\delta_{11},\ldots,\delta_{gn_g}$). This is called right censoring. Since we will consider only the right censoring in this article, we will use "censoring" as a short for "right censoring". Suppose the survival function of $t_{ij}$ based on a postulated model is defined as $S_{ij}(t_{ij}^{*}) = P(t_{ij}> t_{ij}^{*})$, where the subscript $ij$ indicates that the probability depends covariate $x_{ij}$ for the $j$th individual of the $i$th group.

For a shared frailty model, the hazard of an event at time $t$ for the $j$th individual, $j$ = 1, 2, . . . , $n_i$, in the $i$th group, is then
\begin{equation}
h_{ij}(t) =z_i \exp(x_{ij} \beta)h_0(t);
\end{equation}
and the survival function for the $j$th individual of the $i$th group at time $t$ follows:
\begin{equation}
S_{ij}(t) = \exp \bigg\{ -  \int_{0}^{t} h_{ij}(t) \, \mathrm{d}t \bigg \}
=\exp \bigg\{-z_i \exp(x_{ij} \beta) H_0(t) \bigg \},
\end{equation}
where $x_{ij}$ is a row vector of values of $p$ explanatory variables for the $j$th individual in the $i$th group, i.e., $x=(x_{11}, \ldots, x_{gn_g})$; $\beta$ is the vector of regression coefficients; $h_0(t)$ is the baseline hazard function, $H_0(t)$ is the baseline  cumulative hazard function, and $z_i$ is the frailty term that is common for all $n_i$ individuals within the $i$th group. Let $z=(z_1, \ldots, z_g)$. The hazard and survival functions with frailty effect can also be written as,
\begin{equation}
h_{ij}(t) = \exp(x_{ij} \beta + u_i)h_0(t),
\end{equation}
and
\begin{equation}
S_{ij}(t) = \exp \bigg\{ - \exp(x_{ij} \beta + u_i) H_0(t) \bigg \},
\end{equation}
where $u_i$= $\log (z_i)$ is a random effect in the linear component of the proportional hazards model. Note that $z_i$ cannot be negative, but $u_i$ can be any value. If $u_i$ is zero, correspondingly $z_i$ being one, which means the model does not have frailty. The form of the baseline hazard function may be assumed to be unspecified as a semi-parametric model or fully specified to follow a parametric distribution.

In our study, we focus mainly on the shared gamma frailty model, since gamma distribution is the most common distribution for modelling the frailty effect \cite{collett_modelling_2015}. It is easy to obtain a closed-form representation of the observable survival, cumulative density, and hazard functions due to the simplicity of the Laplace transform \cite{BalanTheodorA2020Atof}. The gamma distribution is a two-parameter distribution with a shape parameter $k$ and scale parameter $\theta$. It takes a variety of shapes as $k$ varies: when $k$ = 1, it is identical to the well-known exponential distribution; when $k$ is large, it takes a bell-shaped form reminiscent of a normal distribution; when $k$ is less than one, it takes exponentially shaped and asymptotic to both the vertical and horizontal axes. Under the assumption $k= \frac{1}{\theta}$, the two-parameter gamma distribution turns into a one-parameter distribution. The expected value is one and the variance is equal to $\theta$.

\section{Cross-validatory Z-residual}\label{sec:cvz}

The Z-residual is transformed from  the randomized survival probability (RSP) introduced in  \cite{LiLonghai2021Mdfc}. The key idea of RSP is to replace the survival probability (SP) of a censored failure time with a uniform random number between 0 and the SP of the censored time. RSPs were proved to have a uniform distribution on (0, 1) under the true model with the true generating parameters. The RSP for $y_{ij}$ in a shared frailty model can be then defined as:

\begin{equation}
S_{ij}^{R}(y_{ij}, d_{ij}, U_{ij}) =
\left\{
\begin{array}{rl}
S_{ij}(y_{ij}), & \text{if $y_{ij}$ is uncensored, i.e., $d_{ij}=1$,}\\
U_{ij}\,S_{ij}(y_{ij}), & \text{if $y_{ij}$ is censored, i.e., $d_{ij}=0$,} 
\end{array}
\right. \label{rsp} 
\end{equation}

\noindent where $U_{ij}$ is a uniform random number on $(0, 1)$, and $S_{ij}(\cdot)$ is the postulated survival function for $y_{ij}$ given $x_{ij}$. $S_{ij}^{R}(y_{ij}, \delta_{ij}, U_{ij})$ is a random number between $0$ and $S_{ij}(y_{ij})$ when $y_{ij}$ is censored. Li et al.\cite{LiLonghai2021Mdfc} illustrated and proved that the RSP is uniformly distributed on $(0,1)$ given $x_{ij}$ under the true model. Therefore, they can be transformed into residuals with any desired distribution. It is preferred to transform them with the normal quantile:
\begin{equation}
r_{ij}^{Z}(y_{ij}, d_{ij}, U_{ij})=-\Phi^{-1} (S_{ij}^R(y_{ij}, d_{ij}, U_{ij})),\label{zresid}
\end{equation}
where $\Phi()$ is the cumulative distribution function (CDF) of a standard normal distribution. We refer to the residuals as defined in \eqref{zresid} as Z-residuals.

Cross-validation (CV) is a re-sampling method for assessing the predictive value. Leave-one-out cross-validation (LOOCV) method is the simplest approach in which each observation is left out as a test case. The outcome from the test data is predicted from a model fitted to the remaining data by using the remaining observations. Since LOOCV is time-consuming, k-fold cross-validation (k-fold CV) is widely used. The observations are randomly divided into $k$ folds of approximately equal size, and observations in one fold are predicted from a model fitted with the observations in the other folds (called training data). In our study, LOOCV and 10-fold CV methods will be used to calculate the cross-validatory Z-residuals.

More specifically, for the LOOCV Z-residual, each observation $t_{ij}^{test}$ is left out from the full dataset with $n$ observations. This dataset with each case is considered as test data and the datasets with the remaining cases are considered as the training dataset, which is used for estimating the parameters. Once the shared frailty model has been fitted to the training dataset, a vector of the estimated regression coefficients, $\hat{\beta'}$, and the estimated frailty effects, $\hat{z_i}$, can be obtained. In addition, the Breslow (1972) estimator \cite{1972DoPC, lin_breslow_2007} is employed for estimating the cumulative baseline hazard to get $\hat{H_0}$ based on the training dataset. The predictive survival function $S_{ij} (y_{ij})$ for the observation $y_{ij}^{test}$ of the test case is given by:
\begin{equation}
\hat{S}_{ij}(y_{ij}^{test}) = \exp \{- \hat{z_i} \exp(\hat{\beta'} x_{ij}) \hat{H}_0(y_{ij}^{test}) \}.
\end{equation}
Then, the RSP for the actually observed $t_{ij}$ of the test case is defined as:
\begin{equation}
\hat{S}_{ij}^{R}(t_{ij}^{test}, d_{ij}, U_{ij})=
\left\{
\begin{array}{rl}
\hat{S}_{ij}(t_{ij}^{test}), & \text{if $t_{ij}^{test}$ is uncensored, i.e., $d_{ij}=1$,}\\
U_{ij}\,\hat{S}_{ij}(t_{ij}^{test}), & \text{if $t_{ij}^{test}$ is censored, i.e., $d_{ij}=0$.} 
\end{array}
\right. 
\end{equation}
The Z-residual for $t_{ij}^{test}$ is given by:
\begin{equation}
\hat{z}_{ij}(t_{ij}^{test}, d_{ij}, U_{ij})=-\Phi^{-1} (\hat{S}_{ij}^R(t_{ij}^{test}, d_{ij}, U_{ij})).
\end{equation}
Repeating these steps $n$ times for each observation, we have $n$ different pairs of training and test datasets and a LOOCV predictive Z-residual is computed  for each observation. 

For implementing cross-validation, each cluster and each value of a categorical covariate should appear at least once in both the training and test datasets. If a cluster has only one observation and it is left out as a test case, the cluster cannot appear in the training dataset. We do not calculate the cross-validatory Z-residual for such observations. This is because the information of $\hat{z_i}$ for such a cluster with only one observation cannot be obtained from the training dataset. Similar requirements are enforced for all categorical covariates. The cross-validatory Z-residuals for such observations are set to be NA in our implementation.

The k-fold CV method splits the full dataset into $k$ groups of observations and the groups are of approximately equal size. One group is left out to form the test dataset, and the dataset of the remaining $k-1$ groups is used as the training dataset for fitting the shared gamma frailty model. The estimates, $\hat{\beta'}$, $\hat{z_i}$, and $\hat{H_0}$, can be obtained from the fitted model, and the predictive Z-residuals are calculated for the observations in the test dataset.  These steps are the same as the LOOCV method described above.  In splitting the dataset into $k$ groups, we try to make each group contain a similar number of observations of each cluster and of each category of categorical covariates. In creating the cross-validation folds, we ensure that the set of cluster identities and the values of each categorical covariate in the test dataset is a subset of the corresponding values in the training dataset. In addition, we also avoid the situation that there is no event (observed) failure time for a certain cluster or a certain category of categorical covariates, which is not allowed in fitting non-parametric Cox proportional hazard models with the \texttt{survival} package. 

\section{Simulation Studies and Results}\label{sec:simulation}
\subsection{Detection of Non-linear Covariate Effect} \label{sec:simulation1}

In this section, we compare the performance of Z-residuals with and without cross-validation in detecting non-linear covariate effects via simulation studies. We generate failure times $t_{ij}$ from a Weibull regression model with shape parameter $\alpha$=3 and scale parameter $\lambda$=0.007, as follows: 
\begin{equation}  
t_{ij}= \bigg( \frac {- \log(v_{ij})}{\lambda z_i \exp(x_{ij}^{(1)} +\beta_2 \log(x_{ij}^{(2)}) + 0.5 x_{ij}^{(3)} )} \bigg)^{1 / \alpha },
\end{equation}
\noindent where $i$ = \{1,$\ldots$, 10\} and $j $ = \{1, $\ldots$, $m$ \} and $v_{ij}$ is simulated from Uniform(0, 1).  The censoring times $C_{i}$ is simulated from an exponential distribution, $\exp(\theta)$, where $\theta$ is set to have censoring rates ($c$) approximately equal to $50\%$. The three covariates are generated as follows: $x_{ij}^{(1)}$ from Uniform(0, 1), $x_{ij}^{(2)}$ from positive-Normal(0, 1), and  $x_{ij}^{(3)}$ from Bern(0.25). The frailty term is generated from the gamma distribution with a mean of 1 and a variance of 0.5. 

\begin{figure}[htp]
\centering
\includegraphics[width=1\textwidth, height=0.6\textwidth]{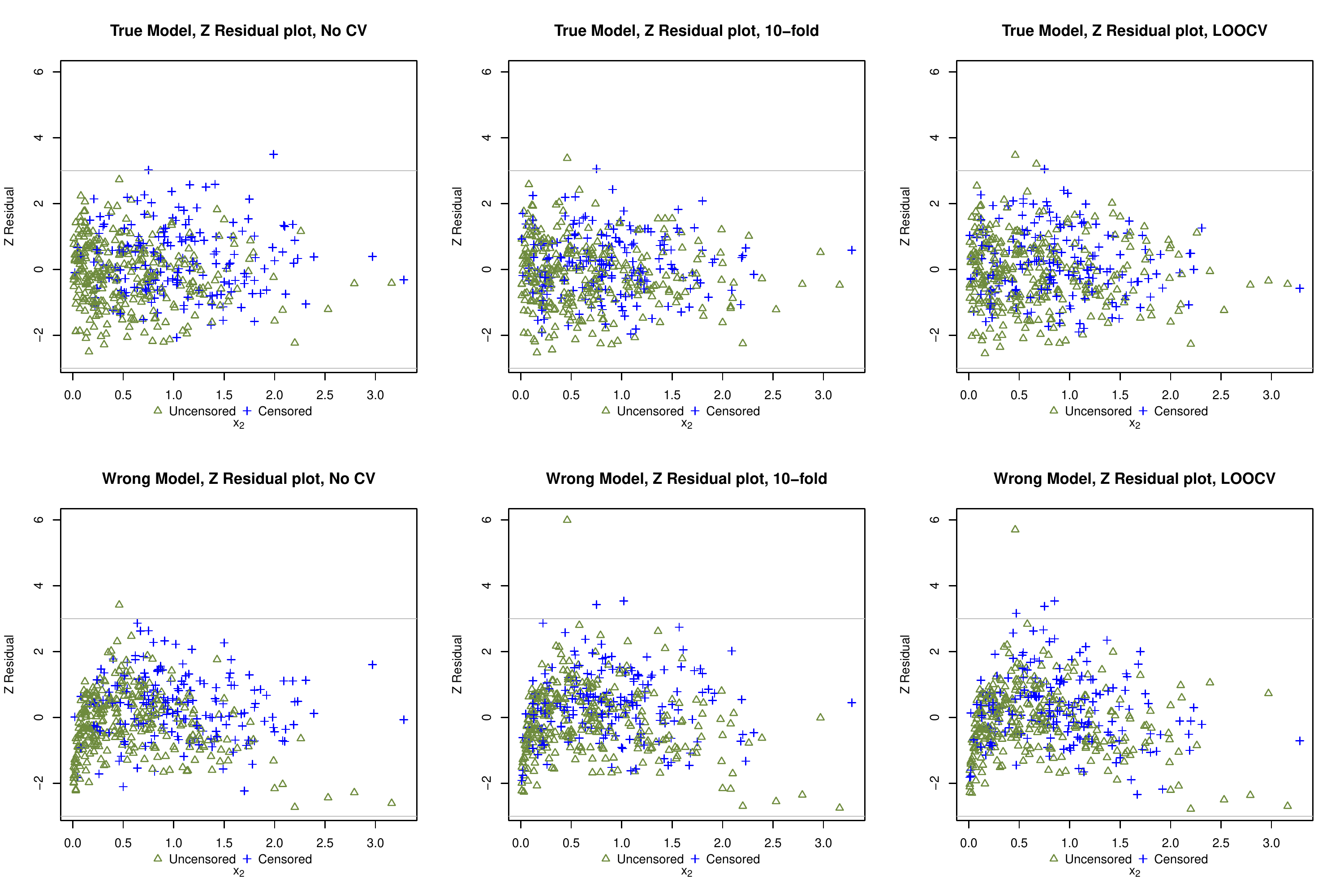}
\caption{The scatterplots of the No-CV, 10-fold and LOOCV Z-residuals for a simulated dataset with non-linear covariate effect, described in Section \ref{sec:simulation1}. The sample size is 500 (10 clusters of 50 observations), the censoring percentage is 50\%, and the $\beta_2$ for $\log(x^{(2)})$ is set to -2. The gray horizontal lines indicate the values 3 and -3. The green points are event times and the blue points are censored times.}
\label{fig:residual_plot_6s}
\end{figure}

We consider fitting a shared frailty gamma model assuming $h_{ij}(t) =z_i \exp(\beta_1 x_{ij}^{(1)} + \beta_2 x_{ij}^{(2)} +\beta_3 x_{ij}^{(3)} )h_0(t)$ as a wrong model, and fitting a shared frailty gamma model assuming $h_{ij}(t) =z_i \exp(\beta_1 x_{ij}^{(1)} + \beta_2 \log(x_{ij}^{(2)}) +\beta_3 x_{ij}^{(3)} )h_0(t)$ as the true model. We first visualize the difference of the Z-residuals with and without cross-validation on a single dataset generated with a strong non-linearity covariate effect ($\beta_2=-2$). The dataset has 10 clusters of 50 observations (the total sample size $n=500$). The scatterplots of Z-residuals against the covariate $ x_{ij}^{(2)}$  are shown in Fig. \ref{fig:residual_plot_6s}, in which the two rows show the true and the wrong models and the three columns show three different methods for computing Z-residuals --- the No-CV, 10-fold and LOOCV methods respectively. Under the true model, the scatterplots of the three types of Z-residuals are randomly scattered without exhibiting any pattern and they are mostly within the interval (-3, 3). Note that most Z-residuals are concentrated on the left side of the x-axis because $x_{ij}^{(2)}$ was simulated from the positive Normal(0, 1). Under the wrong model, all the scatterplots of the three types of Z-residuals show a non-linear pattern. However, we see that, for one observation, the Z-residuals computed by cross-validation methods are near the value 6, but the corresponding No-CV Z-residual is near 3. In addition, there are more cross-validatory Z-residuals greater than 3 than No-CV Z-residuals. This is an indicator of the conservatism of the No-CV method. The QQ plot of the three types of Z-residuals under the true model aligns nearly perfectly with the $45^\circ$ straight line  in the appended Fig. \ref{fig:qq_plot_6s}. Under the wrong model, the QQ plot of the No-CV Z-residuals aligns with the diagonal line; however, the 10-fold and LOOCV Z-residuals show more severe deviations from the $45^\circ$ straight line in the upper tail, demonstrating the increased power of the cross-validatory Z-residuals in detecting non-linear covariate effects compared to the No-CV Z-residuals.

\begin{figure}[htp]
\includegraphics[width=1\textwidth, height=1\textwidth]{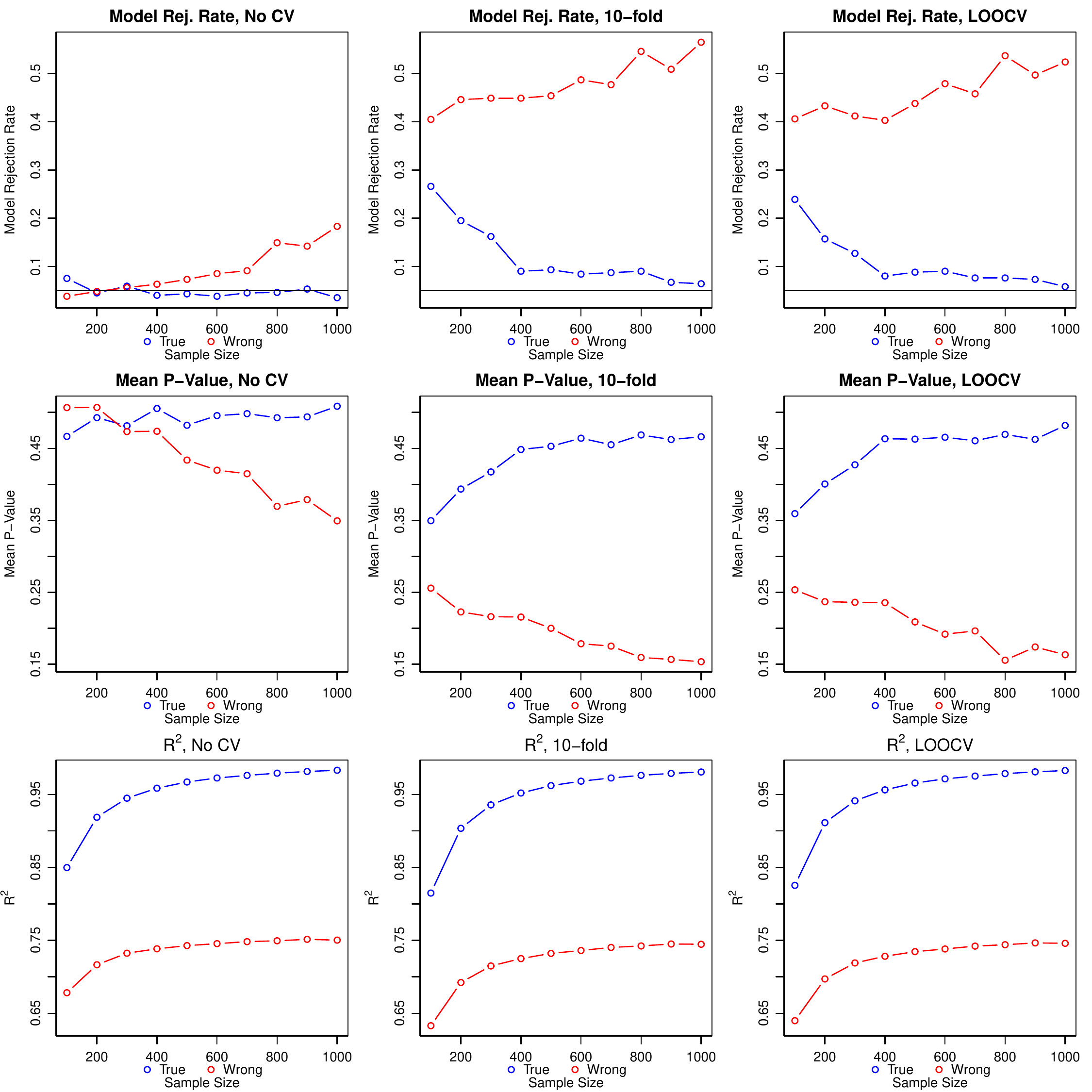}
\caption{Comparison of model rejection rates (proportions of SW test p-values $\leq$ 0.05) and the means of SW p-values with Z-residuals based on the No-CV, 10-fold and LOOCV methods for detecting the non-linear covariate effect. The percentage of censoring is 50\% and the true regression coefficient for the nonlinear covariate, $\log(x_2)$, is -2. The plots in the third row show the values of $R^2$ for measuring the agreement between the survival probabilities calculated with the fitted models and the survival probabilities calculated with the true generating models.  }
\label{fig:low6s}
\end{figure}

We used multiple simulated datasets to investigate the difference between Z-residuals with and without cross-validation when they are used in GOF tests. We apply the Shapiro–Wilk (SW) test to check the normality of the three types of Z-residuals for checking the overall GOF of fitted models. For this investigation, we generated 1000 datasets with 10 clusters of equal size, each having $m$ observations, with $m$ varying in the set of  $\{10,20,\ldots, 100\}$.  We also set two different values for $\beta_2$ (-2 and -1) for representing strong and moderate non-linear covariate effects.  The model rejection rate is estimated by the proportion of SW test p-values less than 0.05 in the 1000 datasets. We also calculated the mean of the SW p-values in the 1000 datasets for comparing the difference with and without cross-validation. Fig. \ref{fig:low6s} presents the results for the scenario with a strong non-linearity effect ($\beta_2=-2$), where the three columns correspond to the No-CV, 10-fold, and LOOCV Z-residuals respectively. The first row of Fig. \ref{fig:low6s} displays the model rejection rates of the SW test under the true (blue lines) and the wrong (red lines) models. Under the true model, the model rejection rates of the No-CV Z-residuals are close to but slightly lower than the nominal level of 0.05 for all scenarios. By contrast, the  SW tests with the 10-fold and LOOCV Z-residuals under the true model have slightly higher type-I error rates than the nominal level of 0.05 (explained below) when the sample size is smaller than 400, but the type-I error rates are close to 0.05 as the sample size increases. More importantly, the powers of the SW tests with the No-CV Z-residuals are very low and significantly lower than the corresponding powers of the SW tests with the 10-fold and LOOCV Z-residuals in all scenarios.  Figure \ref{fig:low6s} also shows that the performances of the  SW tests with the 10-fold and LOOCV Z-residuals are very similar. This finding is practically important because the computation of LOOCV Z-residuals is much more time-consuming. The panels in the second row of Figure \ref{fig:low6s} present the means of SW test p-values. We see that the means of SW p-values of the 10-fold and LOOCV Z-residuals under the wrong models are remarkably smaller than those of the No-CV Z-residuals.  We also observe that the gap between the means of SW p-values under the true and wrong models is significantly larger for cross-validatory Z-residuals than No-CV Z-residuals. In summary, these results suggest that the SW tests with cross-validatory Z-residuals are more powerful in detecting the nonlinear covariate effects than those with No-CV Z-residuals.

We have seen that the SW tests with cross-validatory Z-residuals have slightly larger type-I error rates than the nominal level when the sample size is small (Figure \ref{fig:low6s}). We postulate that the elevation is due to the finite-sample error in estimating the model parameters. In particular, there might be large errors in the estimation of frailties, which can receive information only from the observation within each cluster. In theory, the exact normal distribution for Z-residuals holds  when they are calculated with the true model \textit{with the true parameters.} Given a dataset with a finite sample size, there are still sampling errors in estimating the model parameters,  even though the fitted model has the correctly specified form. As a result, the fitted model is not exactly the true model for the dataset. To illustrate the difference between the fitted and the true models, we calculated the $R^2$ value between the survival probabilities calculated with the parameters estimated with the three different methods (No-CV, 10-fold, LOOCV) and the true survival probabilities calculated with the true generating parameters. The panels in the third row of Figure \ref{fig:low6s} show the average of the $R^2$ values in the 1000 datasets and also in different cross-validation folds for each simulation setting. We see that the $R^2$ for the cases with small sample sizes is substantially smaller than the cases with large sample sizes. Therefore, it is reasonable that the type-I error rates of the SW tests with cross-validatory Z-residuals are larger than the nominal level of 0.05. Figure \ref{fig:low7w} in the appendix provides the results based on the scenario with a moderate non-linear covariate effect ($\beta_2=-1$). The results are generally consistent with the scenario with a strong non-linear covariate effect, but the model rejection rates and means of the SW p-values are slightly lower when the wrong model is fitted to the datasets.

\begin{figure}
\includegraphics[width=0.5\textwidth, height=0.5\textwidth]{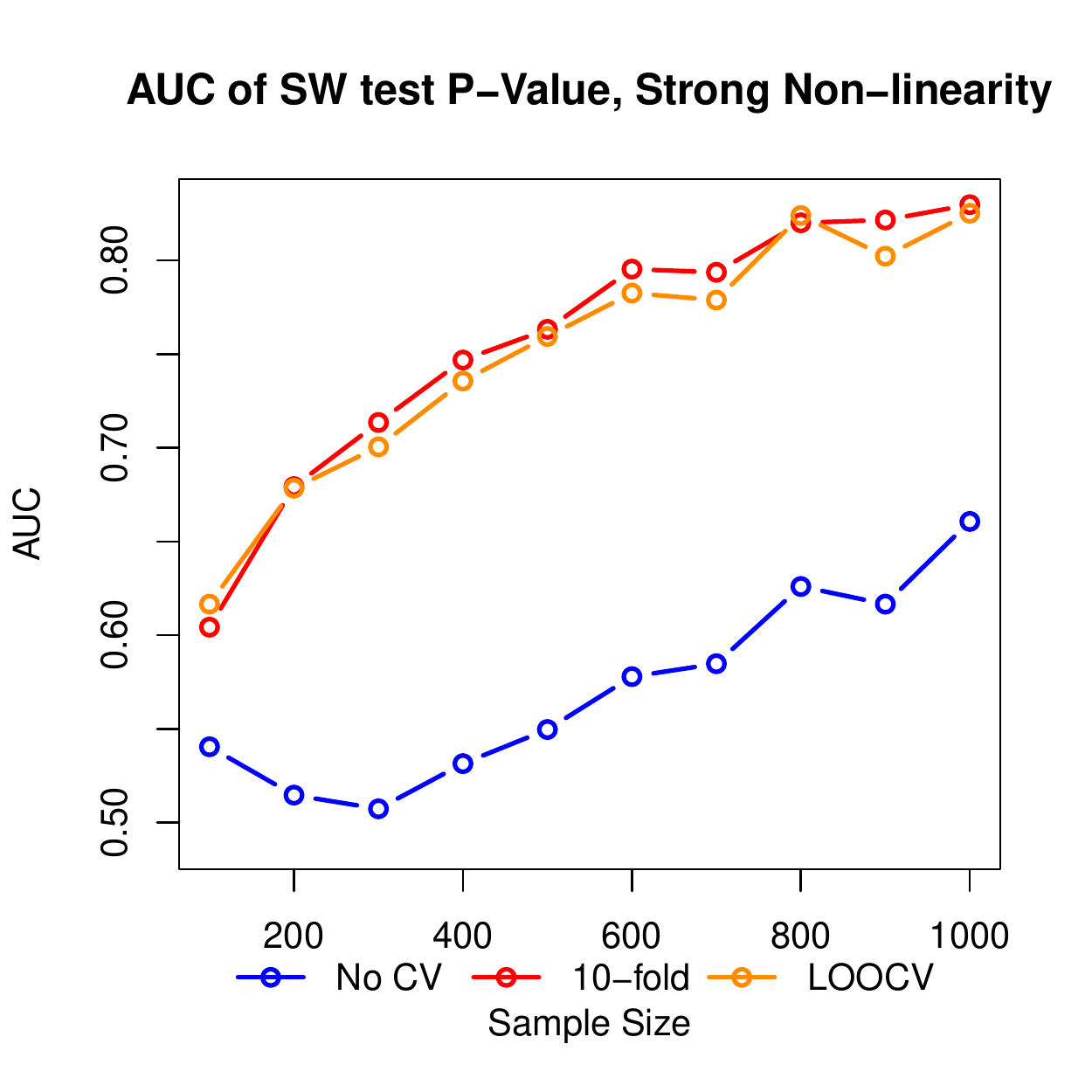}
\includegraphics[width=0.5\textwidth, height=0.5\textwidth]{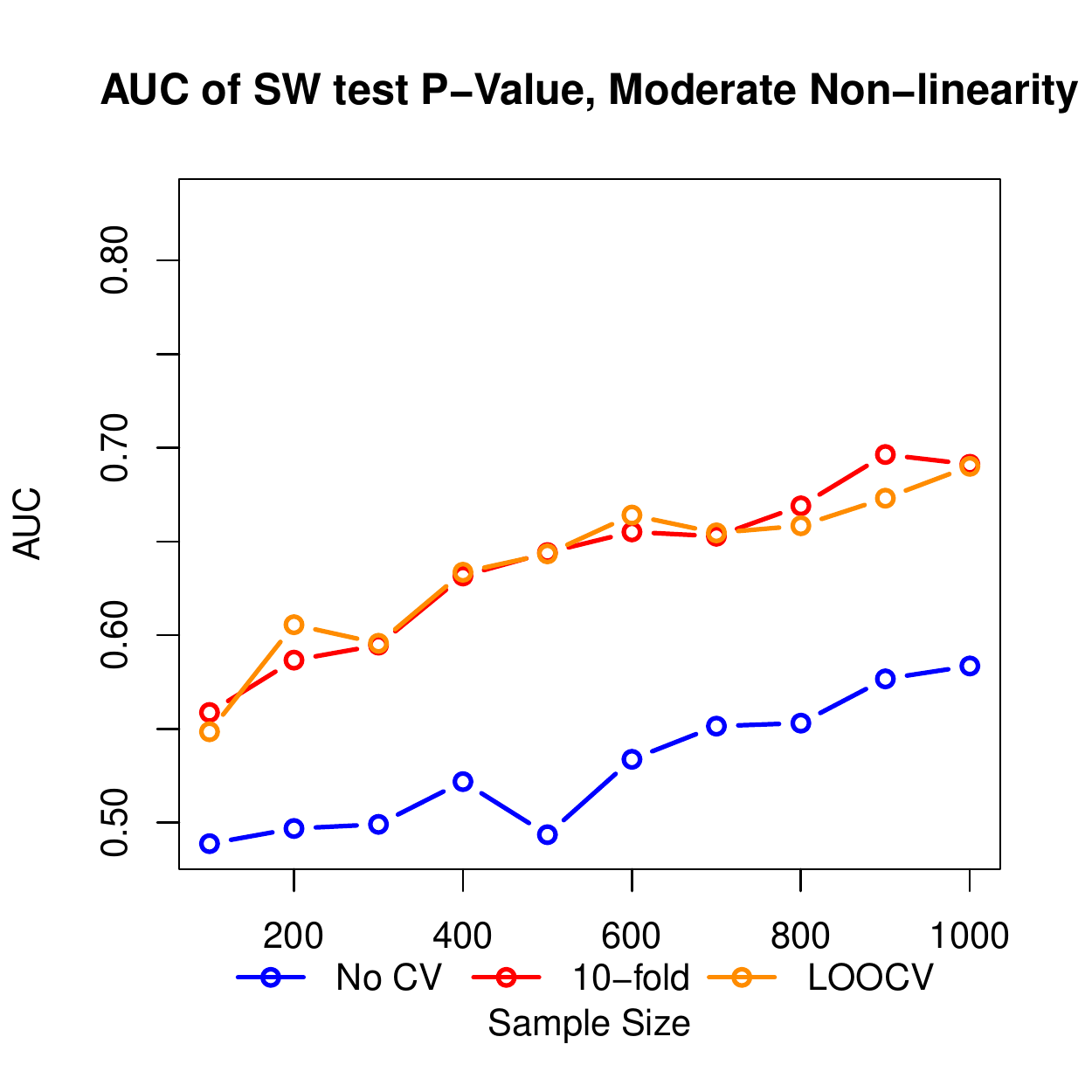}

\caption{Comparison of the AUC values of SW test p-values based on Z-residuals computed with the No-CV, 10-fold and LOOCV methods for simulation datasets with non-linearity effects. }
\label{fig:low_auc67}
\end{figure}

We also use the area under the ROC curve  (AUC) to summarize the discriminative powers of SW test p-values and use them to compare the three types of Z-residuals. The AUC measures the difference between two groups of 1000 SW test p-values, one from fitting the true model and the other from fitting the wrong model. When the AUC is high, the SW p-values are well separated between the two groups, indicating that they are discriminative for discerning adequate and inadequate models. Figure \ref{fig:low_auc67} shows the AUC values for the scenarios with strong and moderate non-linear covariate effects in the left and right plots respectively. The AUC of all three methods increases as the sample size increases, and the AUC values of the 10-fold and LOOCV Z-residuals are very close to each other. More importantly, the AUC values of 10-fold and LOOCV Z-residuals are consistently much higher than the corresponding values of No-CV Z-residuals. Furthermore, we notice that the superiority of cross-validatory Z-residuals does not diminish when the sample size increase, at least up to 1000. This finding is fairly remarkable as we might think that the bias due to the double use of the dataset may disappear when the sample size is sufficiently large. In summary, the SW p-values calculated with cross-validatory Z-residuals are much more discriminative in separating the proper and improper models than those calculated with No-CV Z-residuals. 

\subsection{Detecting Outliers}\label{sec:simulation2}

In this section, we compare the performance of the Z-residuals with and without cross-validation in identifying outliers via simulation studies. We generate a clean dataset from a Weibull model and then add jitters to create a corresponding contaminated dataset, for which we know the identities of outliers. For the clean datasets, we generate the true failure times from a Weibull regression model with shape parameter $\alpha$=3 and scale parameter $\lambda$=0.007 as follows:
\begin{equation}
t_{ij}= \bigg( \frac {- \log(v_{ij})}{\lambda z_i \exp(x_{ij}^{(1)} - 2x_{ij}^{(2)} + 0.5 x_{ij}^{(3)} )} \bigg)^{1 / \alpha }, 
\end{equation}
\noindent where $i$ = \{1,$\ldots$, 10\} and $j $ = \{1, $\ldots$, $m$ \} and $v_{ij}$ is simulated from Uniform (0, 1). The censoring times $C_{ij}$ is simulated from an exponential distribution, $\exp(\theta)$, with $\theta$ being set to obtain censoring rates  approximately equal to $50\%$. The three covariates are generated as follows: $x_{ij}^{(1)}$ from Uniform(0, 1), $x_{ij}^{(2)}$ from Normal(0, 1), and  $x_{ij}^{(3)}$ from Bern(0.25). The frailties are generated from the gamma distribution with a mean of 1 and a variance of 0.5. The jitters added to outliers are generated from $\max(w, e$), where $e$ is a random number from exp(1) and the value $w$ is set to 2 or 4 for indicating moderate and strong jitters respectively. The value $w$ is introduced to ensure that the jitters are at least greater than $w$. We also consider two different schemes of adding jitters to clean datasets. One is adding to randomly selected 10\% event times, and the other is adding to a random selection of 10 event times. Note that the  contaminated failure times may not always appear excessively large if the failure time before contamination is small enough.  We repeatedly simulate 1000 datasets with 10 clusters of $m$ observations, where the cluster size $m$ is varied in the set of \{10, 20, \ldots, 100\} for investigating how the performance of Z-residuals depends on the cluster size. To these simulated datasets, we fit the shared frailty gamma model assuming $h_{ij}(t) =z_i \exp(\beta_1 x_{ij}^{(1)} + \beta_2 x_{ij}^{(2)} +\beta_3 x_{ij}^{(3)} )h_0(t)$, which is the true model generating clean datasets.

\begin{figure}[htpb]
\centering
\includegraphics[width=1\textwidth, height=0.6\textwidth]{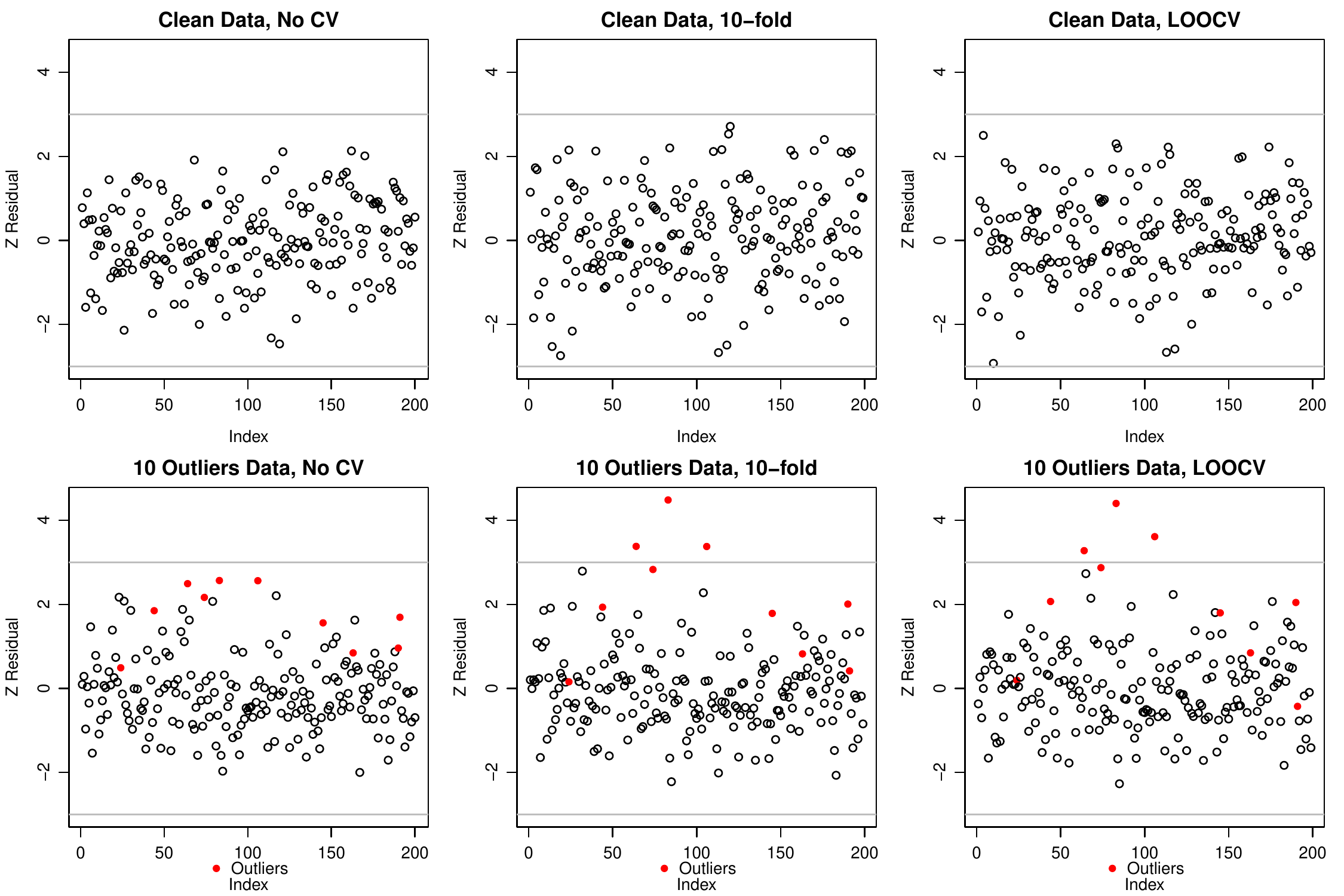}
\caption{Comparison of the performance of the No-CV, 10-fold, and LOOCV Z-residuals in detecting outliers on a pair of clean and contaminated datasets. The datasets have 10 clusters with 20 observations in each.}
\label{fig:outlier_resid}
\end{figure}

We first visualize the difference of Z-residuals with and without cross-validation on a pair of clean and contaminated datasets with the cluster size $m = 20$. Strong jitters are added to 10 randomly selected failure times for generating the corresponding contaminated dataset. Fig. \ref{fig:outlier_resid} displays the residual plots for the clean and contaminated datasets in the first and second rows respectively. The red points indicate the outliers with their failure times being added with jitters. The Z-residuals for the clean dataset are mostly bounded between -3 and 3 as standard normal variates without any unusual patterns. For the contaminated dataset, all the No-CV Z-residuals are bounded between -3 and 3, which means that they fail to detect the outliers if we declare outliers when the Z-residual is out of $(-3,3)$; by contrast,  the cross-validatory Z-residuals of three outliers (red points) fall out of the interval (-3, 3). This comparison suggests that the cross-validatory Z-residuals have increased powers in detecting outliers even though not all the outliers could be detected because their failure times are still not excessive to the model after being added with jitters.

\begin{figure}[htpb]
\includegraphics[width=1\textwidth, height=1\textwidth]
{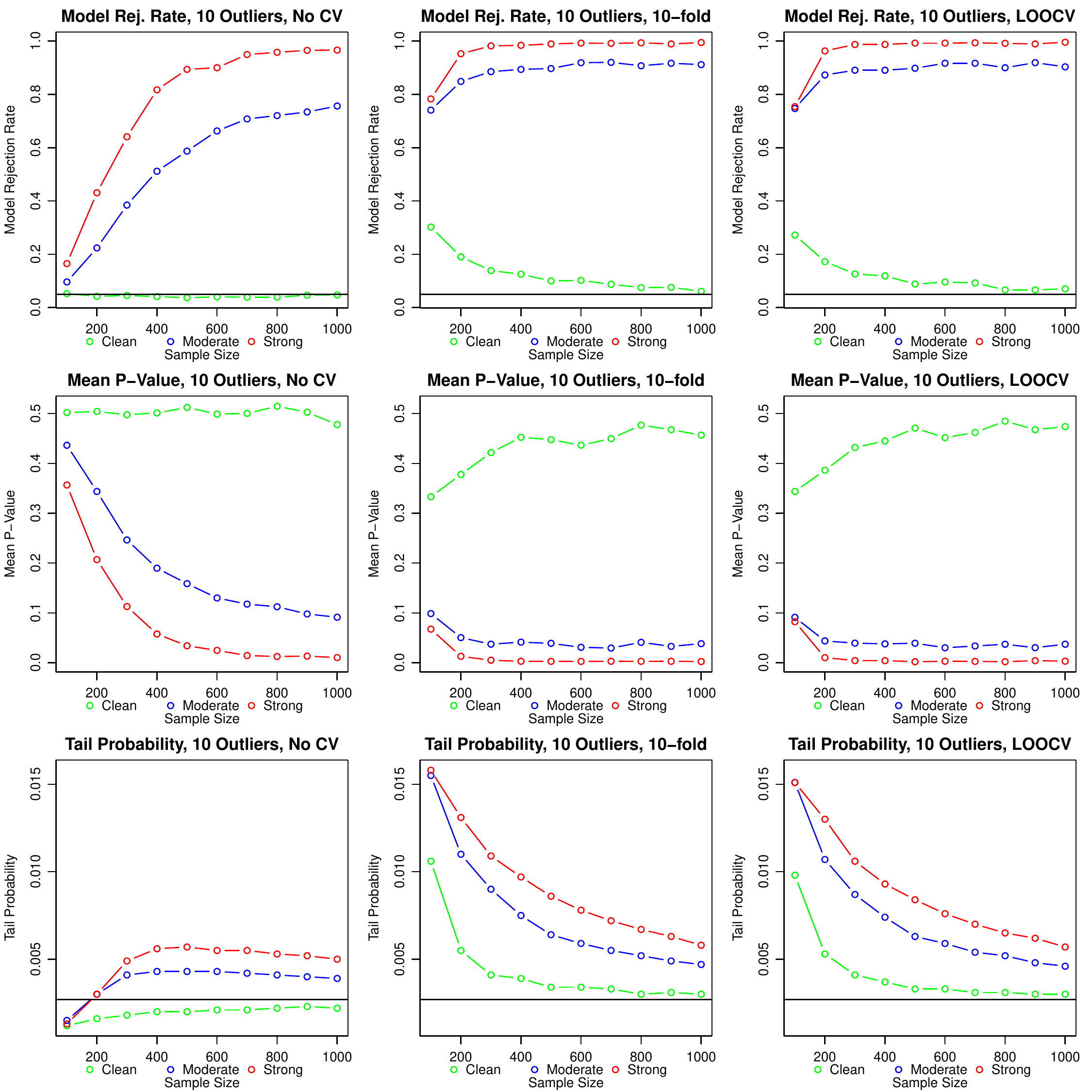}
\caption{Comparison of model rejection rates based on the SW test p-values $\le$ 0.05, the mean of SW p-values,  and the tail probability of the No-CV, 10-fold, and LOOCV Z-residuals for the datasets with 10 outliers. The horizontal lines for the model rejection rate show the nominal type-I error rate of  SW tests under the true model, ie, 0.05. The horizontal lines for the tail probability show the expected value for clean datasets, ie, $P(|Z|>3)=0.0027$ where $Z\sim N(0,1)$.}
\label{fig:outlier1_combm2m4}
\end{figure}

We use multiple simulated datasets to investigate the performance of the  SW tests based on the three types of Z-residuals in identifying model inadequacy for contaminated datasets. We fit the true model for clean datasets to both clean and contaminated datasets. This model is adequate for clean datasets but is inadequate for contaminated datasets, which require a more sophisticated model. Therefore, we expect that the  SW test should have a high chance (power) of rejecting this model for contaminated datasets. As in Section \ref{sec:simulation1}, we use 1000 simulated datasets for each simulation setting to calculate the proportion of the SW test p-values less than 0.05 and calculate the mean of SW p-values. Figure \ref{fig:outlier1_combm2m4} shows the results for the scenario with 10 strong outliers.  As displayed in Figure \ref{fig:outlier1_combm2m4}, the model rejection rates of No-CV Z-residuals for clean datasets (green line) remain at the nominal level of 0.05 for all scenarios. However, for contaminated datasets, the model rejection rates (powers) of No-CV Z-residuals are much lower than the corresponding powers of 10-fold and LOOCV Z-residuals; the reduction in powers is substantial when the sample size is less than 300, for example from about 0.8 to about 0.2 when $m=10$.  We also notice that the type-I error rates (for the clean datasets) of the SW tests with cross-validatory Z-residuals are slightly higher than the nominal level of 0.05 when the sample size is small; nevertheless, they approach 0.05 as the sample size increases.  From the second row of Figure \ref{fig:outlier1_combm2m4}, we also observe that the means of the SW p-values of the No-CV Z-residual are much higher than those of cross-validatory Z-residuals.  For investigating the performance in outlier detection, we also calculate a tail probability about Z-residuals, which is the proportion of Z-residuals with absolute values greater than 3, which is often used to identify outliers in practice. The plots in the third row of Figure \ref{fig:outlier1_combm2m4} show the means of the tail probabilities in 1000 simulated datasets under different simulation scenarios. The tail probabilities of No-CV Z-residuals for clean datasets are all below the expected value of 0.0027, which is $P(|Z|>3)$ where $Z\sim N(0,1)$.  More importantly, we see that the tail probabilities of 10-fold and LOOCV Z-residuals for the contaminated datasets are much higher than those of No-CV Z-residuals, and the tail probabilities of 10-fold and LOOCV Z-residuals for clean datasets converge to the expected tail probability --- 0.0027. Figure \ref{fig:outlier2_combm2m4} in the Appendix displays the results for the scenarios in which contaminated datasets have 10\% outliers, and the results are consistent with the scenarios with 10 outliers.

\begin{figure}[htpb]
  \centering
   \begin{subfigure}{0.35\textwidth}
    \centering
    \includegraphics[width=\textwidth, height=\textwidth]{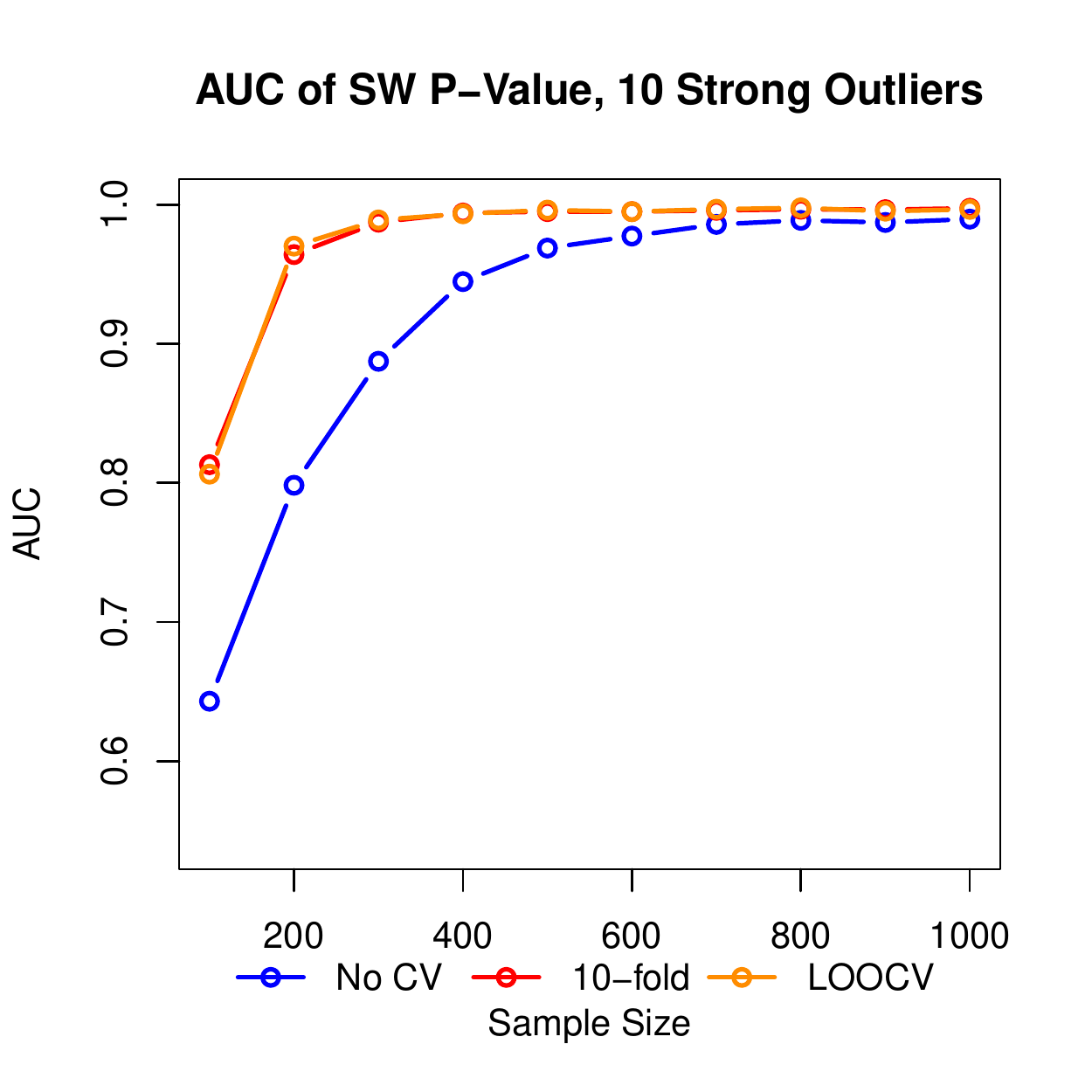}
  \end{subfigure} 
  \begin{subfigure}{0.35\textwidth}
    \centering
    \includegraphics[width=\textwidth, height=\textwidth]{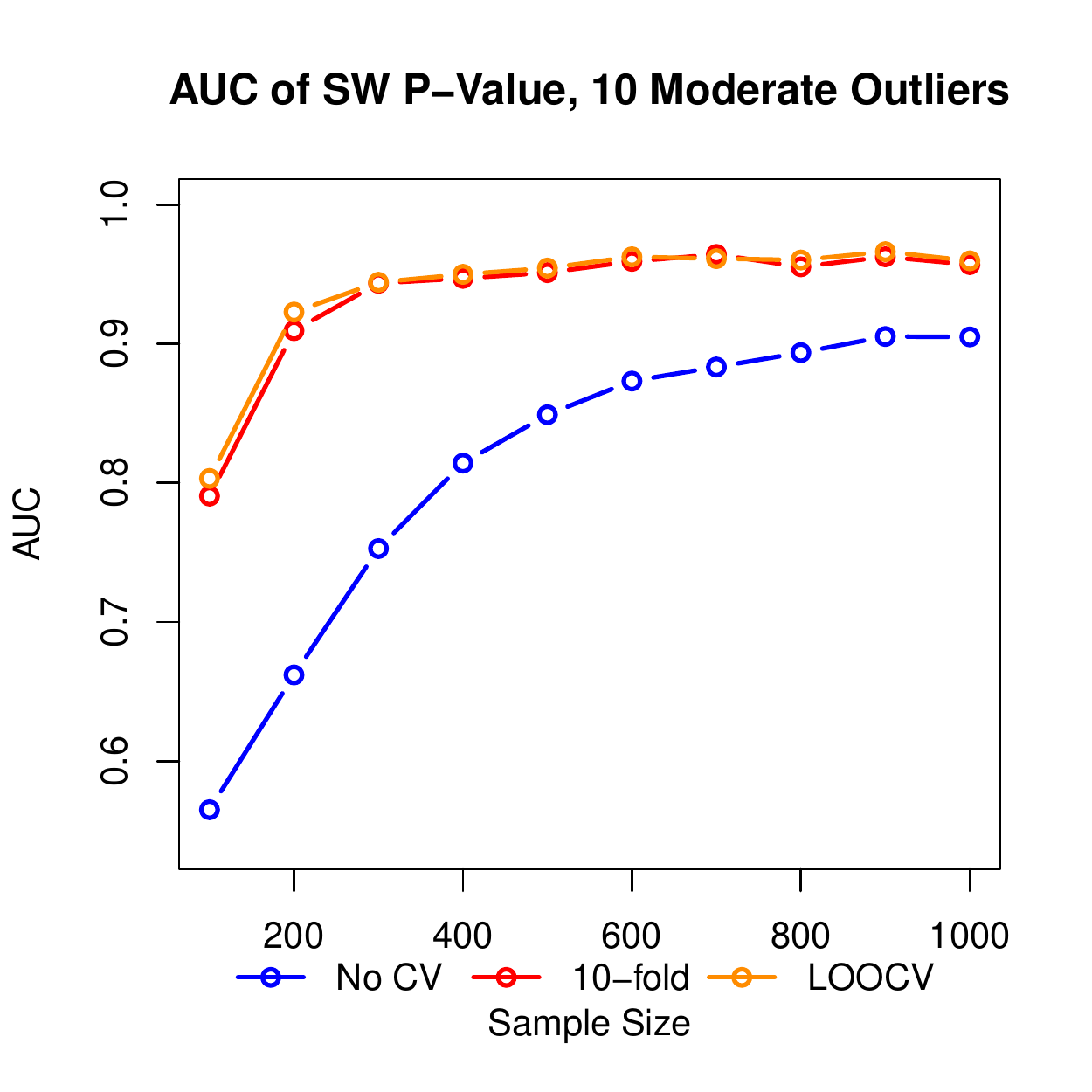}
  \end{subfigure}  

  \begin{subfigure}{0.35\textwidth}
    \centering
    \includegraphics[width=\textwidth, height=\textwidth]{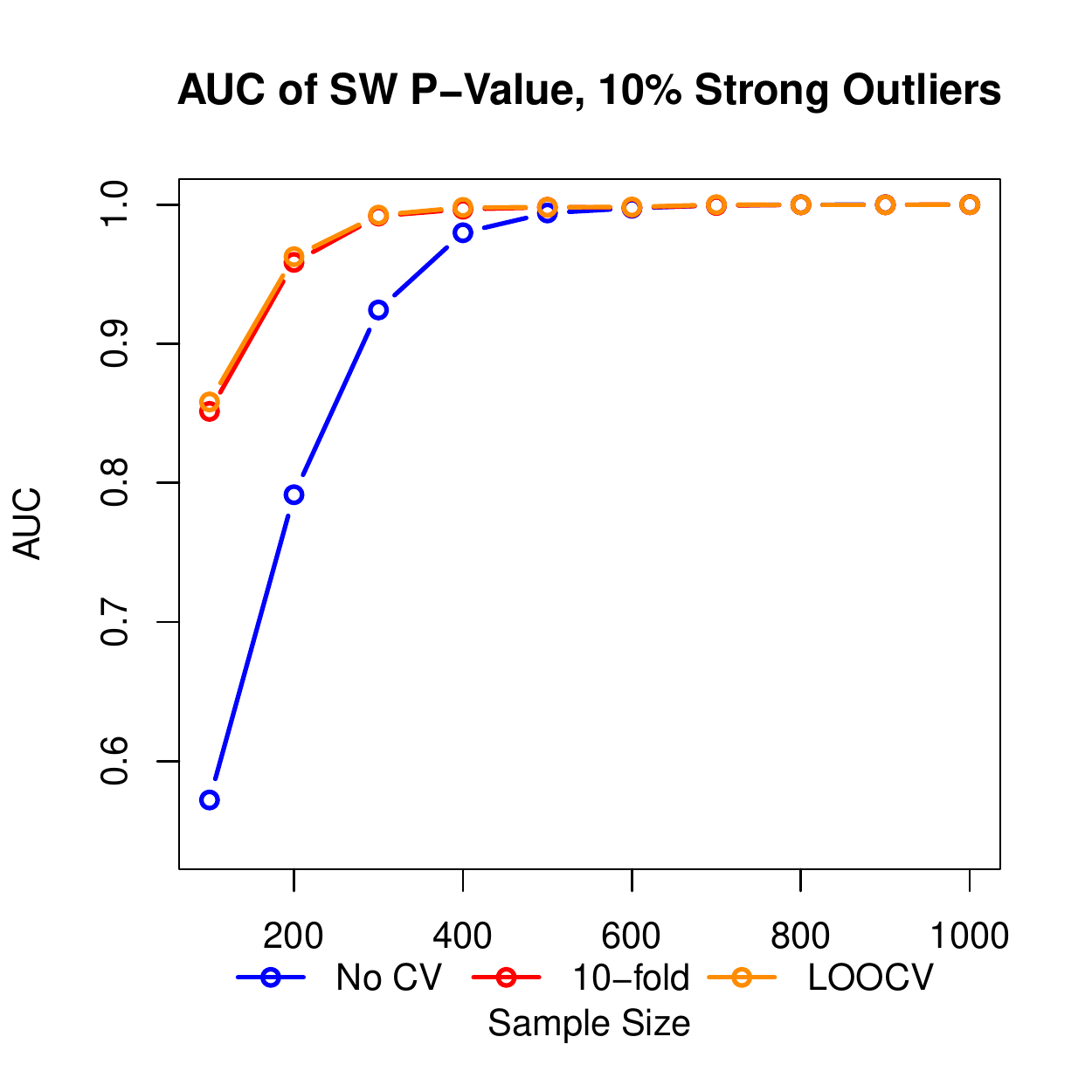}
  \end{subfigure}   
  \begin{subfigure}{0.35\textwidth}
    \centering
    \includegraphics[width=\textwidth, height=\textwidth]{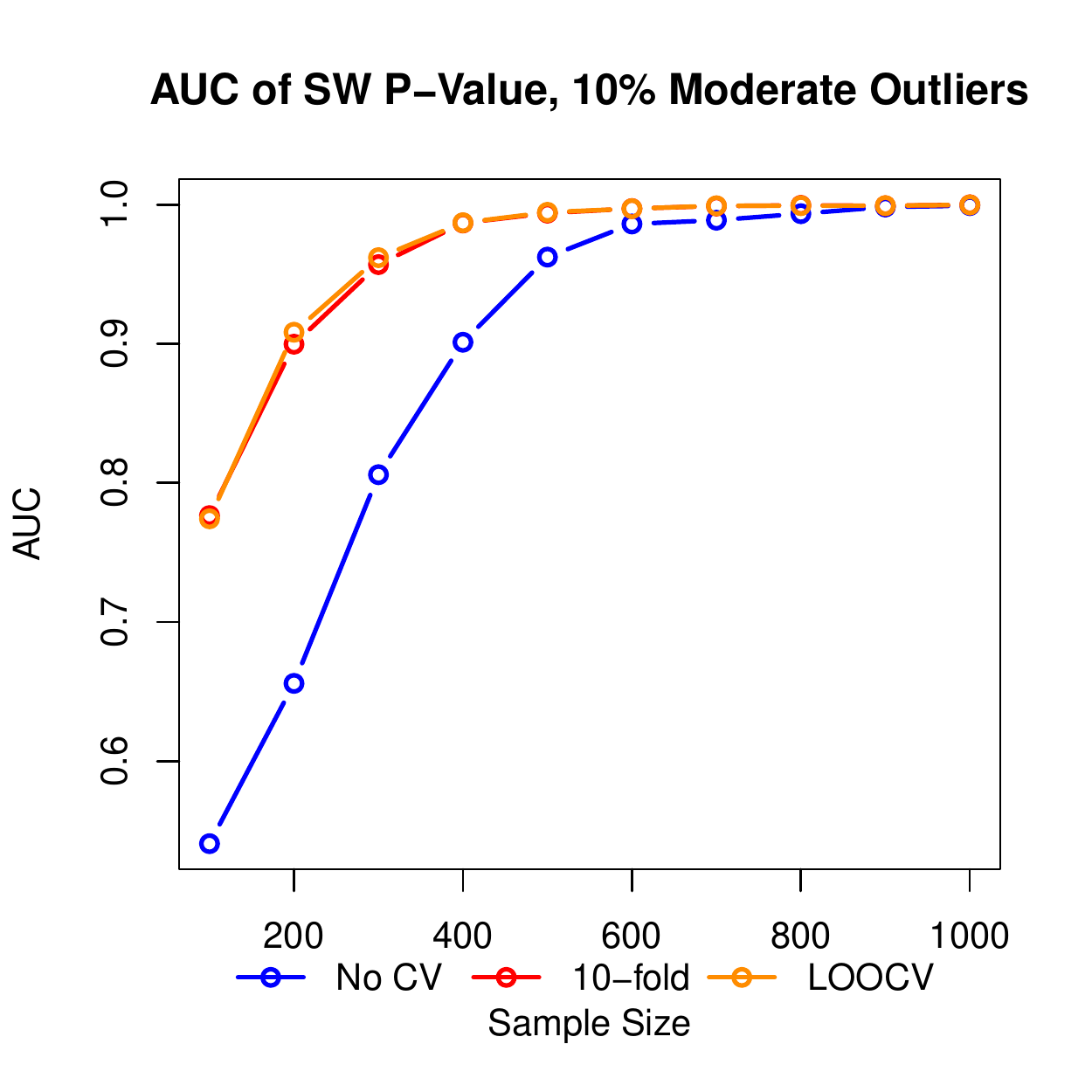}
  \end{subfigure}  
  
\caption{Comparison of the AUC values of SW test p-values based on Z-residuals computed with the No-CV, 10-fold and LOOCV methods for simulation datasets with outliers.} \label{fig: auc_outlier}
 \end{figure}

In order to evaluate the discriminative abilities of SW test p-values, we use the area under the curve (AUC) to measure the difference of the SW test p-values between two groups - one from clean datasets and the other from contaminated datasets. The results are presented in the four plots (Fig. \ref{fig: auc_outlier}) that correspond to four simulation scenarios (combinations of two different levels of jitters and two different schemes for adding jitters). Across all scenarios, we observe that the AUC values of 10-fold and LOOCV Z-residuals are significantly higher than those of No-CV Z-residuals. When the sample size is around 100, the AUC values of No-CV Z-residuals are near 0.5, indicating no discriminative power, whereas the corresponding AUC values of 10-fold and LOOCV Z-residuals are approximately 0.8. Additionally, we notice that the difference in AUC values between Z-residuals with and without cross-validation diminishes to 0 as the sample size increases for three of the four scenarios. However, in the scenario where the number of moderate outliers is fixed at 10, the gap remains visible  even when the sample size is 1000.

\begin{figure}[htp]
  \centering

   \begin{subfigure}{0.35\textwidth}
    \centering
   \includegraphics[width=\textwidth, height=\textwidth]{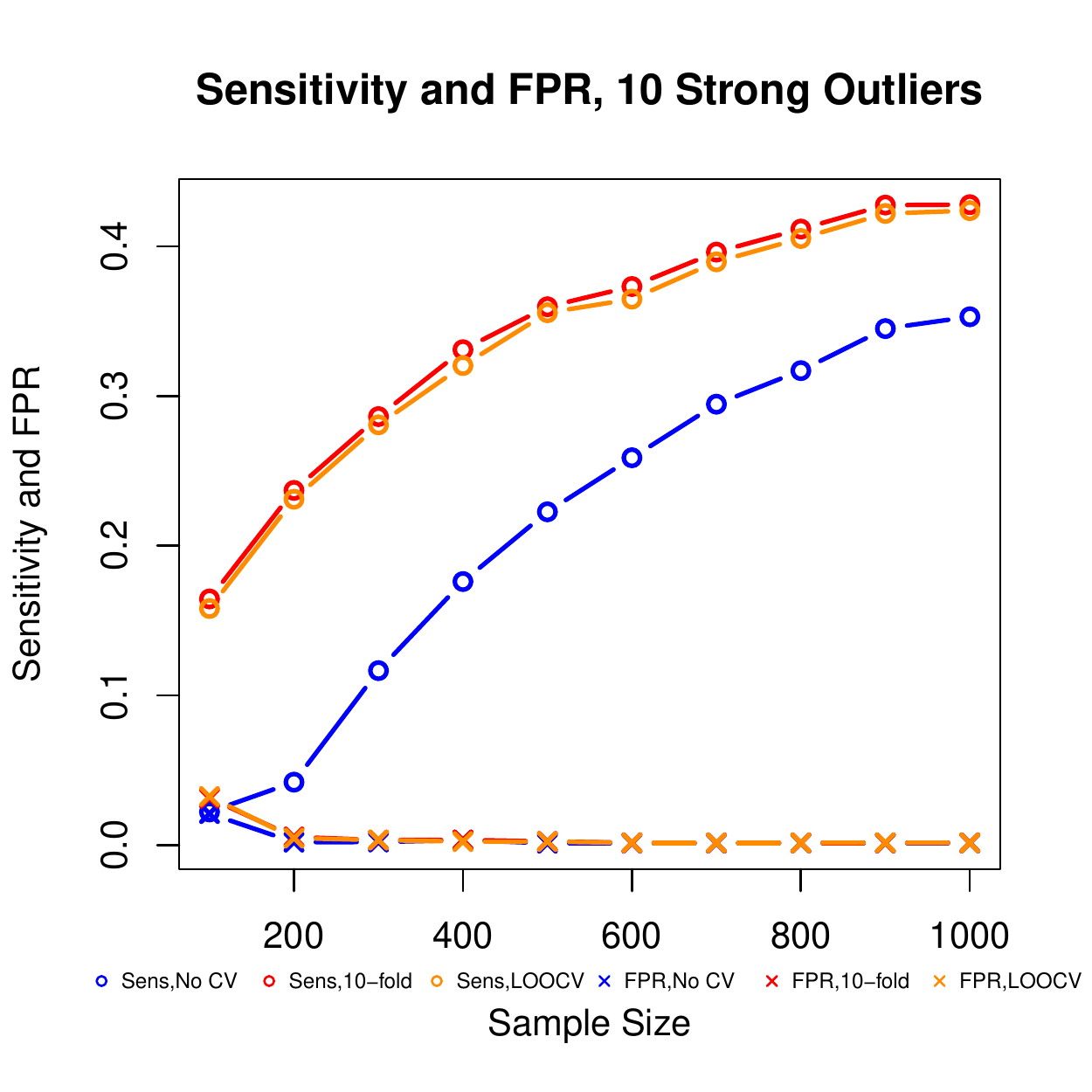}
     \caption{ \label{fig:sens_fpr_10outlierm4}}
  \end{subfigure} 
  \begin{subfigure}{0.35\textwidth}
    \centering
  \includegraphics[width=\textwidth, height=\textwidth]{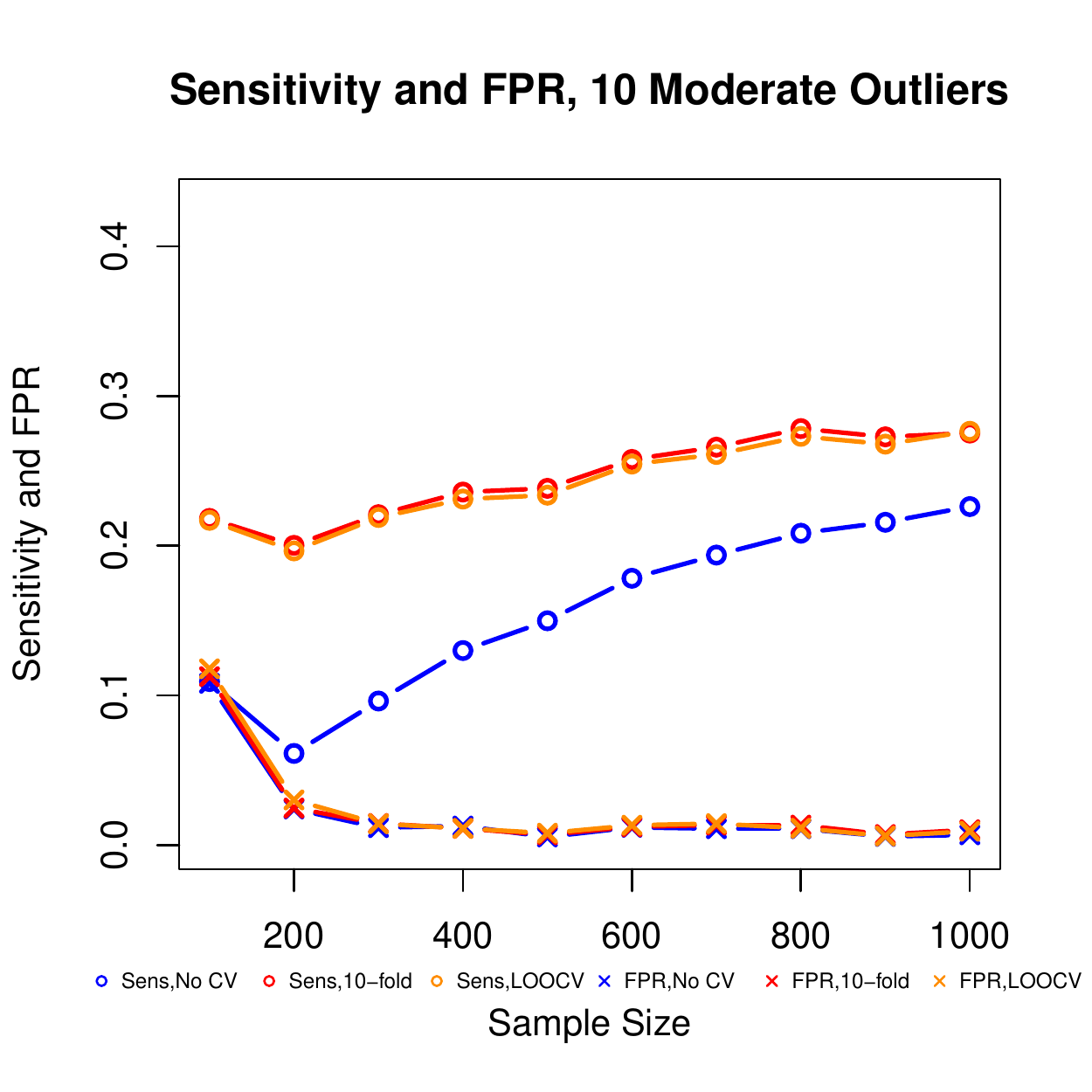}
    \caption{ \label{fig:sens_fpr_10outlier_m2}}
  \end{subfigure}  

  \begin{subfigure}{0.35\textwidth}
    \centering
   \includegraphics[width=\textwidth, height=\textwidth]{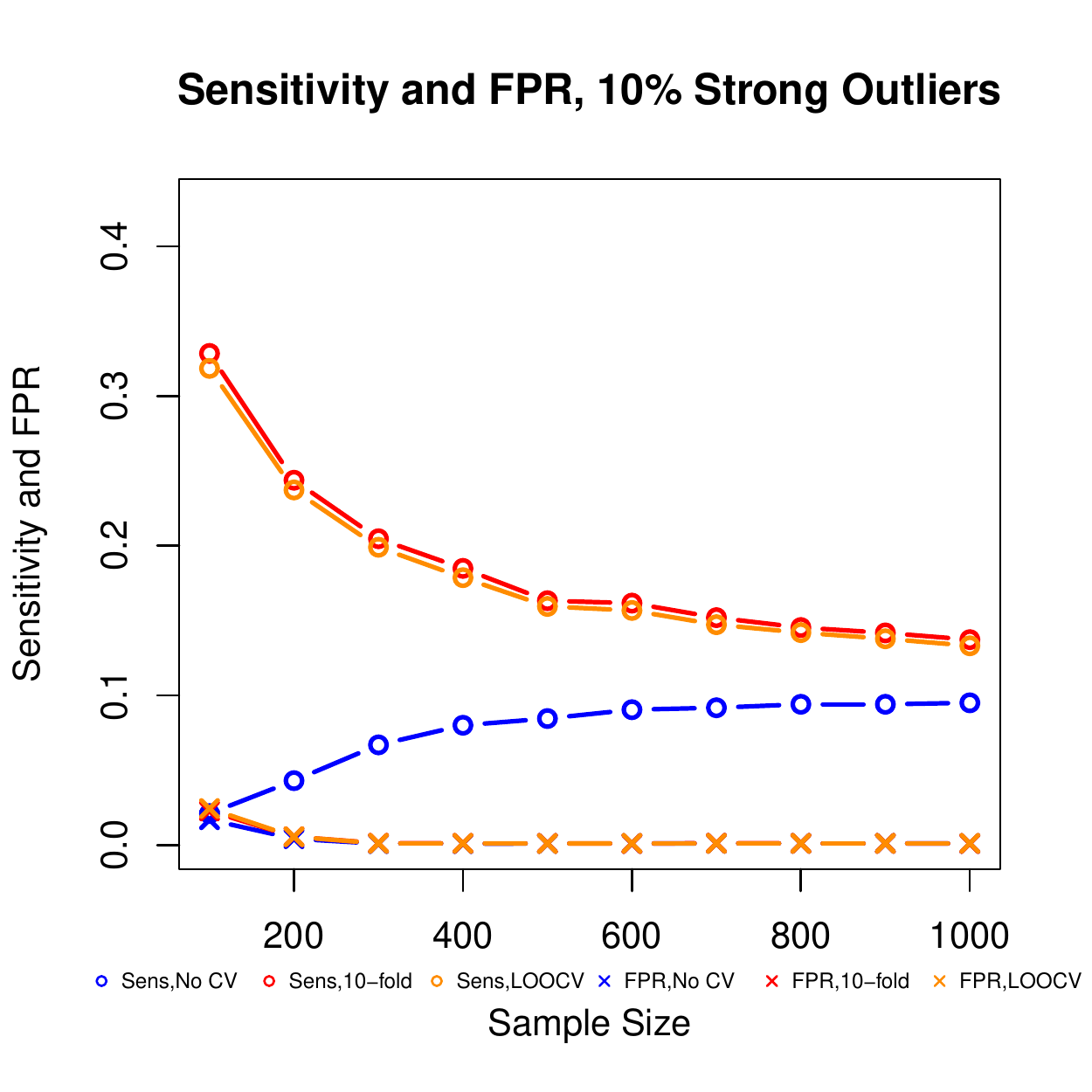}
    \caption{ \label{fig:sens_fpr_outlierm4}}
  \end{subfigure}   
  \begin{subfigure}{0.35\textwidth}
    \centering
   \includegraphics[width=\textwidth, height=\textwidth]{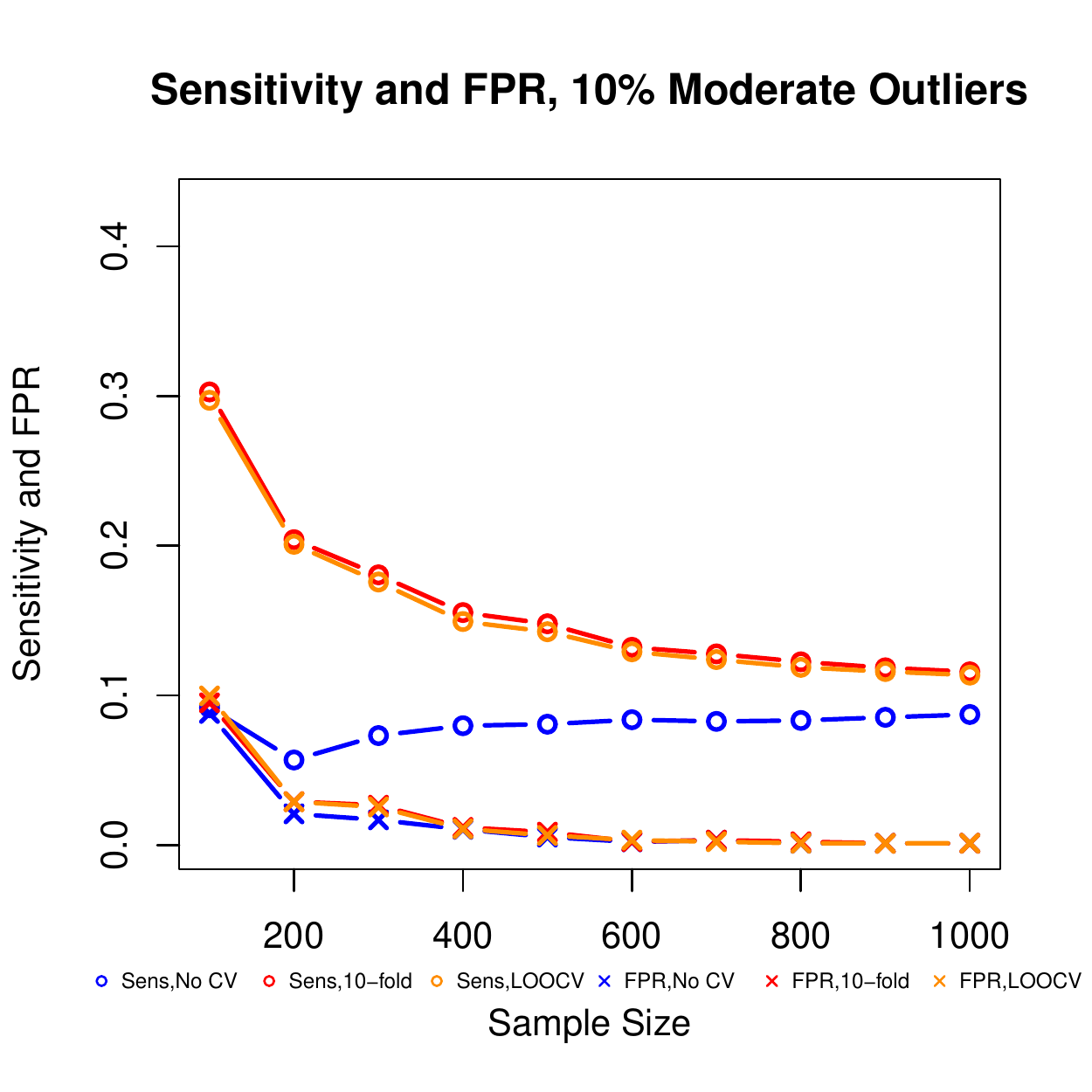}
    \caption{ }\label{fig:sens_fpr_outlierm2}
  \end{subfigure}  

\caption{ Comparison of the sensitivities (points with $\circ$) and the false positive rates (points with $\times$)  in detecting outliers using No-CV, 10-fold, and LOOCV Z-residuals.\label{fig:sens_fpr_outlier}}
 \end{figure}

Finally, we compare the sensitivity and false positive rate (FPR) in detecting outliers using Z-residuals with and without cross-validation. Our rule for identifying an outlier is that the absolute value of its Z-residual is greater than 3. Given this rule, the sensitivity is the proportion of the true outliers that are correctly identified as outliers, and the FPR is the proportion of non-outliers that are falsely identified as outliers. Figure \ref{fig:sens_fpr_outlier} shows the sensitivities and FPRs for the four simulation scenarios as we consider for Figure \ref{fig: auc_outlier}. Clearly, we see that 10-fold and LOOCV Z-residuals have much higher sensitivities and almost the same FPRs when they are used to detect true outliers compared to the No-CV Z-residuals, for all the considered four scenarios. This comparison demonstrates clearly the advantage of using cross-validatory Z-residuals for the purpose of identifying outliers, although we previously see that the SW tests based on cross-validatory Z-residuals have a slight elevation of type-I error rates. Interestingly, we see that the sensitivity of 10-fold and LOOCV Z-residuals increases as the sample size increase when the number of outliers is fixed at 10, but it decreases and converges to a value of about 0.1 when the percentage of outliers is fixed at 10\%.

\section{A Real Data Example}\label{sec:realdata}
In this section, we will demonstrate the effectiveness of the cross-validatory Z-residuals in identifying outliers in a real data application studying kidney infection \cite{MCGILCHRISTC.A1991RwFi}. The dataset consists of 38 kidney patients using a portable dialysis machine, and the times of the first and second recurrences of the kidney infection are recorded for these patients. Each survival time is defined as the time until infection since the insertion of the catheter. The same patient is considered as a cluster because of shared frailty describing the common patient’s effect. If a catheter is removed for reasons other than infection, the observation is considered censored. The censoring percentage is 24\%. The dataset contains 38 patients (cluster), and each patient has exactly two observations, with a total sample size of 76. This data has often been used to illustrate a shared frailty model. More details on this dataset can be found from \cite{MCGILCHRISTC.A1991RwFi}. 

We fit a linear shared gamma frailty model with three covariates --- age in year, sex of male or female, and four different disease types (0=GN, 1=AN, 2=PKD, 3=Other) to the recurrence failure times, with details given in Table \ref{tab:variable} in the Appendix. The fitting is done with the \texttt{coxph} function in the \texttt{survival} package. Table \ref{tab:covariates1} shows the estimated regression coefficients, the corresponding standard errors, and p-values for the covariate effects from fitting the shared gamma frailty model with the full dataset.  The results shown in Table \ref{tab:covariates1} indicate that the two covariates, sex and disease type of PKD, are significantly associated with the hazard of recurrence of kidney infection.

\begin{table}[!htb]
    \caption{Parameter estimates of three shared gamma frailty models fitted with the kidney infection dataset. The tables  \eqref{tab:covariates2} and \eqref{tab:covariates3} show the estimates for two subsets of the original datasets with two and three cases removed as they are identified as outliers with LOOCV Z-residuals. \label{tab:covariates}}
            \scriptsize
    \begin{subtable}{.33\linewidth}
     \centering
       \scriptsize
        \caption{The original dataset \label{tab:covariates1}}
\begin{tabular}{|l|l|l|l|}
\hline
Covariate&$\hat \beta$ &SE & p-value \\
\hline
Age & 0.003&0.011  & 0.775  \\
\hline
Sex:Male & 1.480 &0.358&0.000  \\
\hline
D:GN &0.088 &0.406 &0.829 \\
\hline
D:AN & 0.351 &0.400&0.380 \\
\hline
D:PKD &-1.430 &0.631 &0.023 \\
\hline
Frailty &  &  &0.933 \\
\hline
\end{tabular}
   \end{subtable}%
   \hfill
      \begin{subtable}{.33\linewidth}
      \centering
        \scriptsize
        \caption{Exclusing two outliers \label{tab:covariates2}}
        \begin{tabular}{| l|l|l|l|}
        \hline
Covariate&$\hat \beta$ &SE & p-value \\
\hline
Age & 0.007 &0.011 & 0.530  \\
\hline
Sex:Male & 2.117 &0.400&0.000  \\
\hline
D:GN &0.359 &0.406 &0.380 \\
\hline
D:AN & 0.349&0.407&0.390\\
\hline
D:PKD &-0.797 &0.638 &0.210 \\
\hline
Frailty &  &  &0.940 \\
\hline
\end{tabular}
   \end{subtable} %
\hfill 
     \begin{subtable}{.33\linewidth}
     \centering
        \scriptsize
        \caption{Excluding three outliers \label{tab:covariates3}}
        \begin{tabular}{| l |l | l | l |}
\hline
Covariate&$\hat \beta$ &SE & p-value \\
\hline
Age & 0.012 &0.011 & 0.280  \\
\hline
Sex:Male & 2.120 &0.402&0.000  \\
\hline
D:GN &0.727 &0.415 &0.080 \\
\hline
D:AN & 0.319&0.404&0.430\\
\hline
D:PKD &-0.802 &0.636 &0.210 \\
\hline
Frailty &  &  &0.940 \\
\hline
\end{tabular}
   \end{subtable} 
   
\end{table}

\begin{figure}[htpb]
\includegraphics[width=1\textwidth, height=0.5\textwidth]{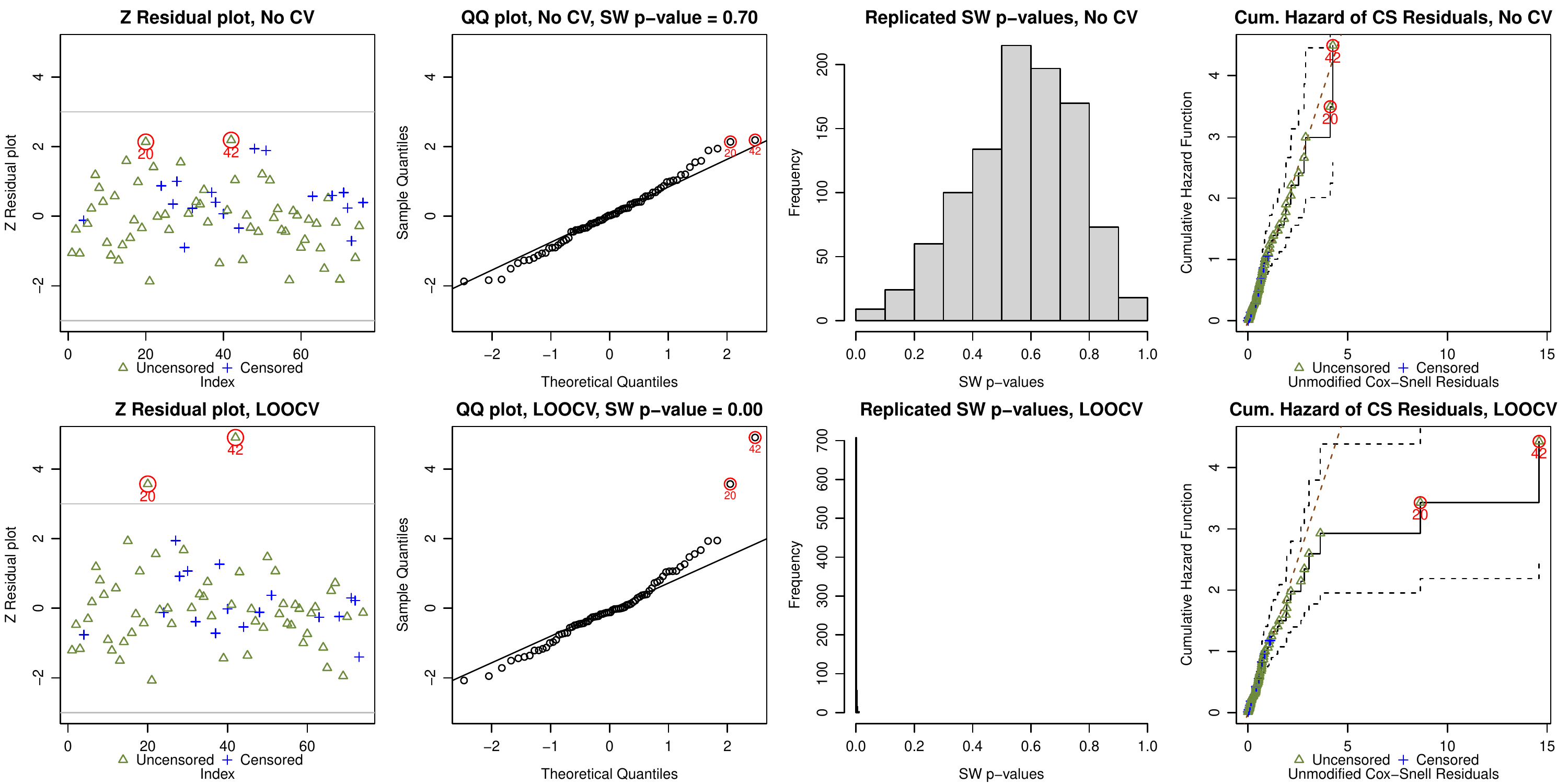}
\caption{Scatterplots and QQ plots of No-CV and LOOCV Z-residuals of the fitted shared frailty models based on the original kidney infection dataset. The third column presents the histograms of 1000 replicated SW p-values of Z-residuals The fourth column shows the CS residuals computed with the No-CV and LOOCV methods.}
\label{fig:kidney1}
\end{figure}

We calculated Z-residuals and Cox Snell (CS) residuals using the No-CV and LOOCV methods for this dataset. We only considered the LOOCV  method since the 10-fold CV and LOOCV Z-residual methods perform very similarly as shown in the simulation studies and the computational burden to implement LOOCV is not a concern due to the small sample size. Figure \ref{fig:kidney1} shows the residual diagnosis results for the original kidney infection dataset without removing outliers. The first and second columns of Figure \ref{fig:kidney1} present the scatterplots versus the index and the QQ plots of the Z-residuals computed with the No-CV and LOOCV methods. The No-CV Z-residuals are mostly between -3 and 3 and the QQ plot of NO-CV Z-residuals aligns well with the $45^\circ$ straight line. The SW p-value of No-CV Z-residuals is about 0.70 as shown in the QQ plot. Such a large p-value indicates a good fit of the model to the dataset. In summary, the diagnosis results with No-CV Z-residuals suggest that the shared frailty model appears appropriate for the dataset and no outlier is identified.  However, the scatterplot of LOOCV Z-residuals shows that the Z-residuals of the two cases labelled with numbers 20 and 42 are greater than 3. We can consider these two cases as outliers for the shared frailty model. The QQ plot of LOOCV Z-residuals shows a large deviation from the $45^\circ$ straight line, which is clearly caused by the large Z-residuals of the two outliers. The SW p-value of LOOCV Z-residuals is also very small --- less than 0.01, as shown in the QQ plot. In summary, the diagnosis results with LOOCV Z-residuals suggest that the fitted shared frailty model is inadequate for this dataset and there are two cases with excessive Z-residuals, which are identified as outliers for this model.  

Compared to all the raw infection times as shown in the appended Figure \ref{fig:kidney_timeplot}, the infection time of case 42 is the highest value among all but does not appear very outstanding; the infection time of case 20 is near the median of all infection times, hence, does not appear outlying at all. This observation illustrates the difference between the concepts of outliers relative to raw observations and relative to a fitted model. Z-residual is a monotone transformation of the tail (or survival) probabilities of the conditional distribution of failure time given covariates (see equation \eqref{zresid}). Therefore, the identification of outliers based on Z-residuals has considered covariate effects. However, the identification of outliers by examining only raw failure times does not consider covariate effects; in other words, it is based on a model with only the intercept term.

There is randomness in the Z-residuals of the censored observations. For the same dataset, we can produce different sets of Z-residuals with different random numbers. Therefore, we would like to replicate a large number of realizations of Z-residuals to see the robustness of the above diagnosis.  The third column of Figure \ref{fig:kidney1} displays the histograms of 1000 SW test p-values, each given by a set of No-CV or LOOCV Z-residuals. The histograms show that more than 95\% of the SW p-values of No-CV Z-residuals are larger than 0.05; however, 100\% of the SW p-values of LOOCV Z-residuals are smaller than 0.05. Therefore, the judgment of the mis-specification of the shared frailty model for the dataset is not incidental based on a particular set of LOOCV Z-residuals but a consistent conclusion under large-scale replications of Z-residuals. 

To further verify the above diagnosis results and illustrate the effect of cross-validation in residual diagnostics,  we also compute CS residuals with both No-CV and LOOCV methods and plot their cumulative hazard functions (CHFs) in the fourth column of Figure \ref{fig:kidney1}. The CHF of No-CV CS residuals aligns well with the $45^\circ$ straight line, indicating a good model fit for the dataset. However, the CHF of the LOOCV CS residuals deviates from the $45^\circ$ straight line in the upper tail, suggesting the inadequacy of the fitted model to the dataset. The conclusion for checking the model adequacy with CS residuals is consistent with the diagnosis results with Z-residuals. Nevertheless, we notice that the diagnosis with Z-residuals provides more information regarding the nature of the discrepancy of the inadequate model ---  the existence of outliers, as well as a quantitative measure of the statistical significance of the model departure.

Finally, we consider deleting the two outliers (cases 42 and 20) from the original kidney infection dataset and then re-fit the linear shared gamma frailty model. Table \ref{tab:covariates2} shows the covariate DiseasePKD is no longer statistically significant at the level of 5\%, and the effect size for the covariate sex becomes larger. The differences in Table \ref{tab:covariates1} and \ref{tab:covariates2} indicate that parameter estimation and inference may be greatly affected by including outliers, which highlights the importance of model diagnosis and outlier detection in practical data analysis. The appended Figure \ref{fig:kidney2} presents the results of residual diagnosis after excluding these two outliers. The LOOCV Z-residual diagnosis results indicate that the refitted model is a  fairly good model for the dataset without cases 42 and 20. Nevertheless, we notice that case 15 has a Z-residual marginally greater than 3. Although case 15 may not be of great concern, as most of the SW p-values of LOOCV Z-residuals show are greater than 0.05,  we refit the model after further removing case 15. Table \ref{tab:covariates3} shows the parameter estimates based on the kidney infection dataset after excluding the three cases, which are similar to those in Table \ref{tab:covariates2}. The Z-residual diagnosis, as shown in appended Figure \ref{fig:kidney3}, neither suggests evidence that the model fitted with the three cases removed is inadequate for the dataset nor identifies an outlier for the model. 

\section{Conclusions and Discussions}\label{sec:conc}

Residual diagnosis plays a critical role in the model-building process for validating the correctness of a fitted model. However, residuals are typically calculated based on the model fitted to the full dataset, without using a strategy to split the dataset into different subsets for model fitting and validation. The double use of the dataset for model fitting and validation might lead to conservatism in model diagnosis, leading to the reduced power of detecting inadequate model fit and identifying outliers for the model. To the best of our knowledge, cross-validation is rarely used in residual diagnosis for survival analysis. In this paper, we developed cross-validation methods to compute Z-residuals for detecting model inadequacy and identifying outliers in the context of shared frailty models. We compare the performance of cross-validatory (10-fold and LOOCV) Z-residuals and No-CV Z-residuals for the purpose of the overall GOF test and outlier detection. Our simulation studies and the application to a real dataset demonstrate that the residual diagnosis without cross-validation tends to be conservative for detecting model misspecification due to the double use of the data, and the cross-validation methods can improve the power of the SW-test with Z-residuals in detecting model inadequacy and improve the power of Z-residuals in identifying outliers.    

Our simulation studies also reveal that the cross-validation may cause a slight elevation of type-I error rates in SW tests with Z-residuals. As we explained with the $R^2$ between the survival probabilities calculated with the fitted models and the survival probabilities calculated with the true generating models, the elevation might be caused by inaccuracy in estimating the parameter, in particular, the estimation of the frailties in small cluster size situations. For such situations, the model fitting algorithms for shared frailty models could improve on estimating the frailties, for example, with a stronger penalization for the frailties.  Alternatively, another direction is to work on improving the methods for computing the cross-validatory Z-residuals or the methods for conducting SW tests with Z-residuals, with the goal of obtaining Z-residuals that are less aggressive in rejecting models.  Marginalizing the frailties when we calculate the randomized survival probability may be a solution. If we marginalize the frailties, the cluster size may have a smaller impact on the computation of Z-residuals; moreover, the restriction that the size of each cluster must be greater than 1 could be resolved. A comparison of the methods for computing Z-residuals with or without marginalizing the frailties is an interesting topic for future work. Lastly, in the present study, we focused on investigating the performance of cross-validatory Z-residuals in diagnosing shared frailty models; however, the proposed cross-validatory residuals could be more broadly applied in diagnosing other types of regression models.

\appendix
\section{Additional Figures and Tables} 
\renewcommand{\thesection}{A}

\subsection{Supplementary Figures for Section \ref{sec:simulation1}} \label{sec:supp_nonlinear}

\begin{figure}[htpb]
\includegraphics[width=0.9\textwidth, height=0.6\textwidth]{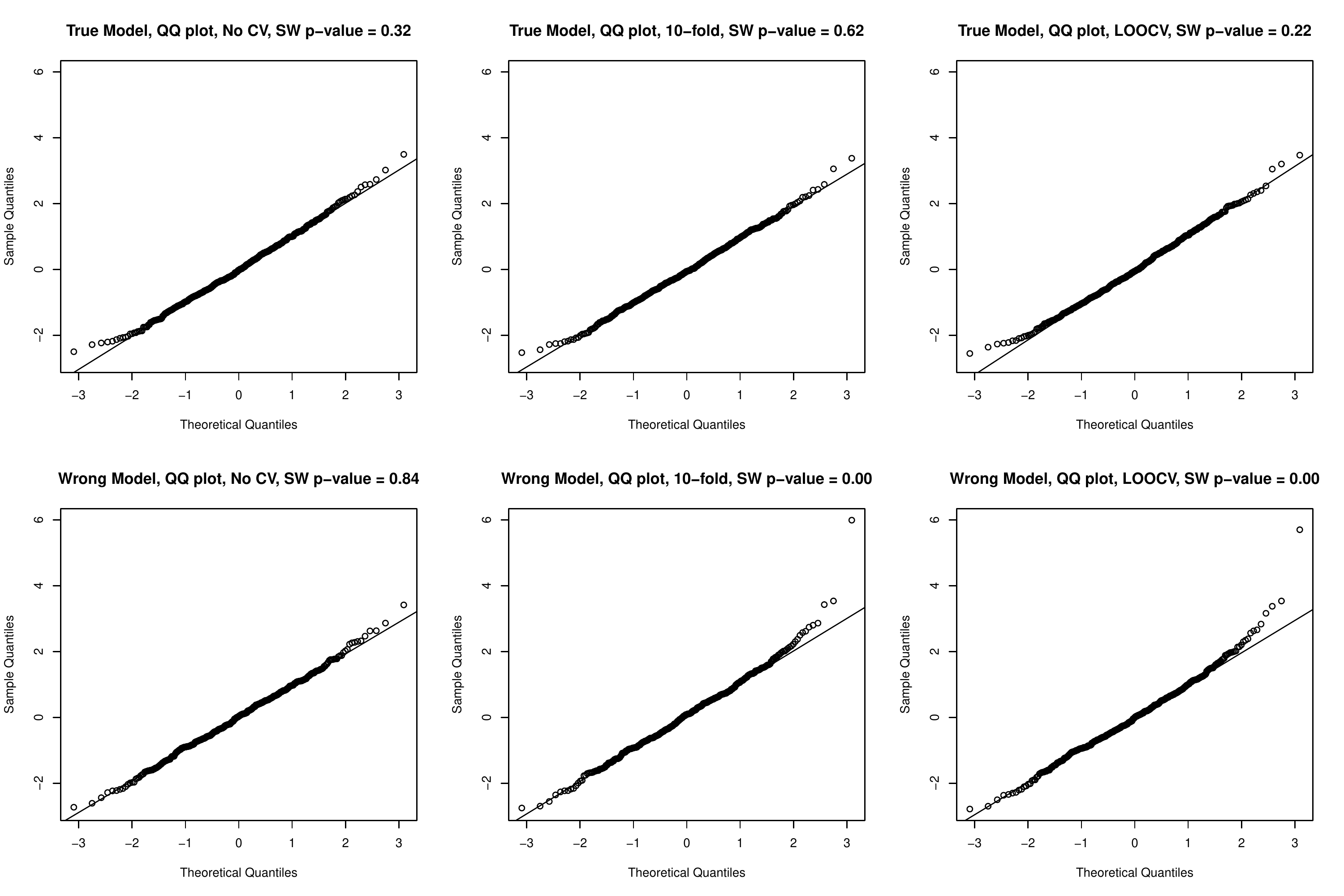}
\caption{The QQ plot of the No-CV, 10-fold and LOOCV Z-residuals as a graphical tool for detecting non-linear effect in covariate with the strong non-linear association. The sample size is 500 (10 clusters of 50 observations), and the censoring percentage is 50\%. }
\label{fig:qq_plot_6s}
\end{figure}

\begin{figure}[htpb]
\includegraphics[width=1\textwidth, height=1\textwidth]{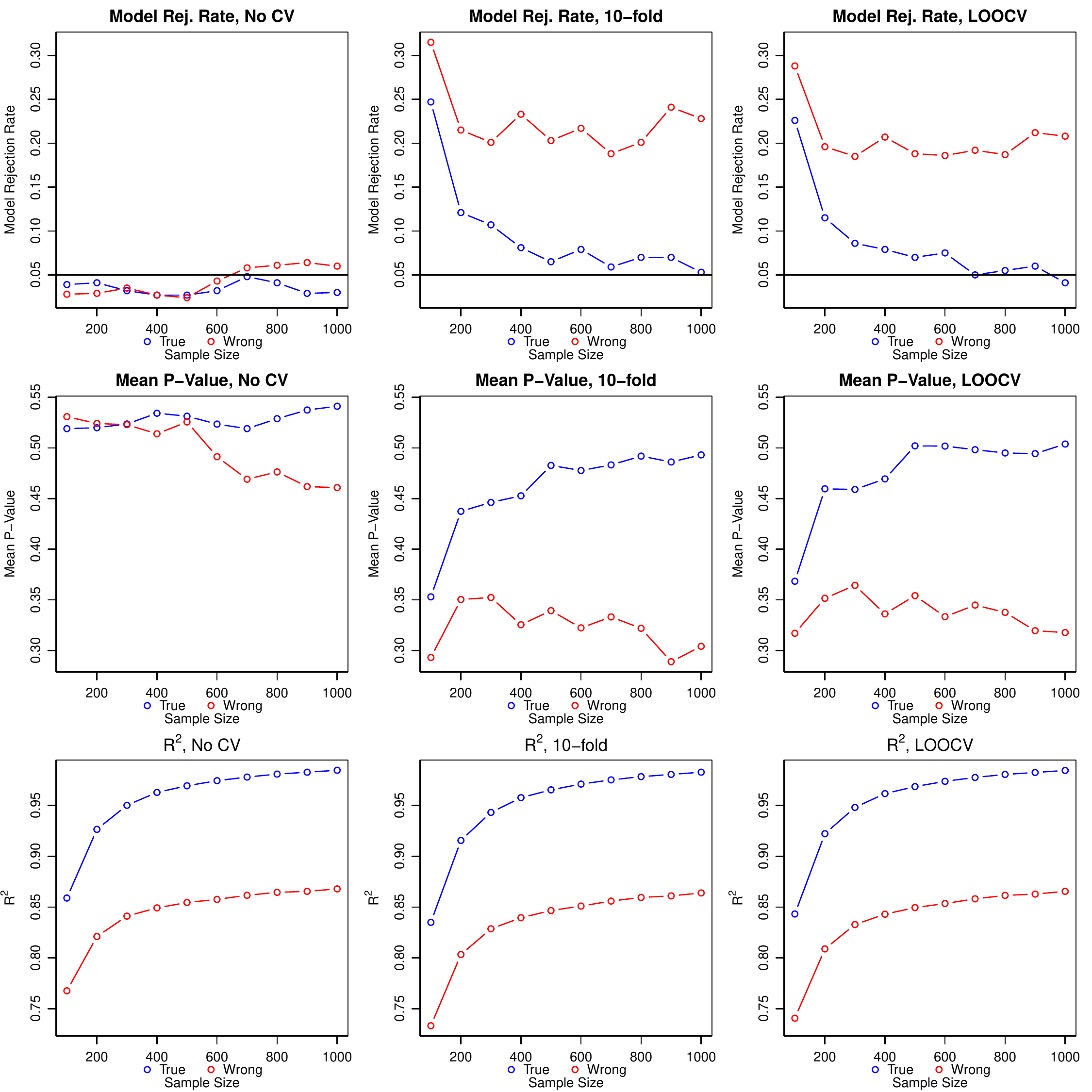}
\caption{Comparison of model rejections based on SW test, the mean of SW p-values and $R^2$ of the No-CV, 10-fold and LOOCV Z-residuals for detecting the moderate non-linear covariate effect.}
\label{fig:low7w}
\end{figure}

\newpage

\subsection{Supplementary Figures for Section \ref{sec:simulation2}}\label{sec:supp_outlier}

\begin{figure}[htpb]
\includegraphics[width=1\textwidth, height=0.9\textwidth]{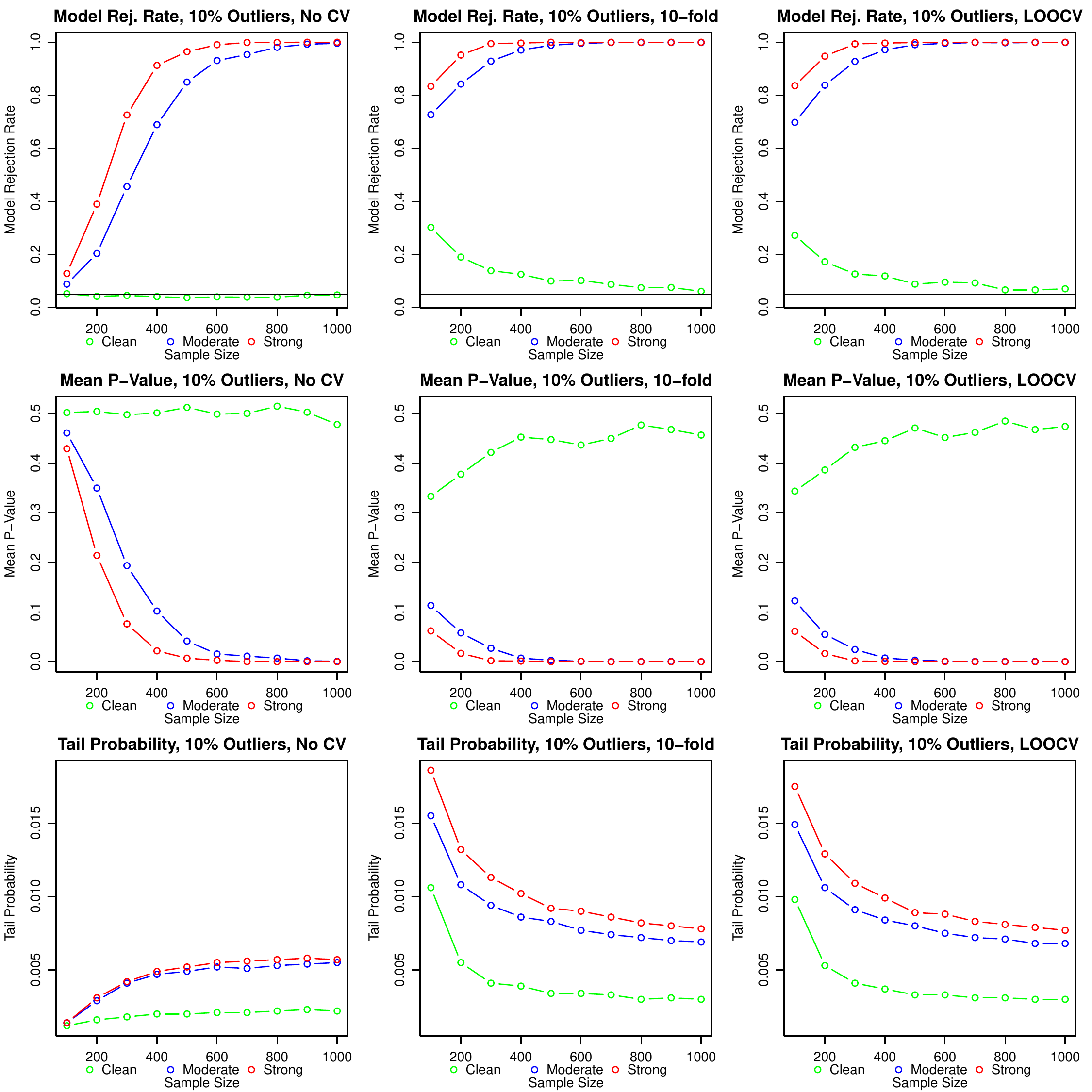}
\caption{Comparison of model rejections rate based on the SW test, the mean of SW p-values and tail probability of the No-CV, 10-fold and LOOCV Z-residuals when the data are contaminated by adding 10\% outliers with moderate and strong deviation from the clean data, respectively.}
\label{fig:outlier2_combm2m4}
\end{figure}

\newpage

\subsection{Supplementary Figures and Tables for Section \ref{sec:realdata}}

\subsection{Details of Variables}

\begin{table}[htpb]
\centering
\caption{Variable definitions for the kidney infection dataset. \label{tab:variable}}
\begin{tabular}{l|l}
\hline
Variable & Definition \\
\hline
$ID$ & Patient number \\
$Time$ & Recurrence time (days) \\
$Status$ & Event indicator (1 = infection occurs; 0 = censored) \\
$Age$ & Patient age (years) \\
$Sex$  & Sex status (1 = male; 2 = female) \\
$Disease$   &Disease Type (0 = GN; 1 = AN; 2 = PKD; 3 = other)\\

\hline
\end{tabular}
\end{table}

\begin{figure}[htpb]
\centering
\includegraphics[width=0.5\textwidth, height=0.5\textwidth]{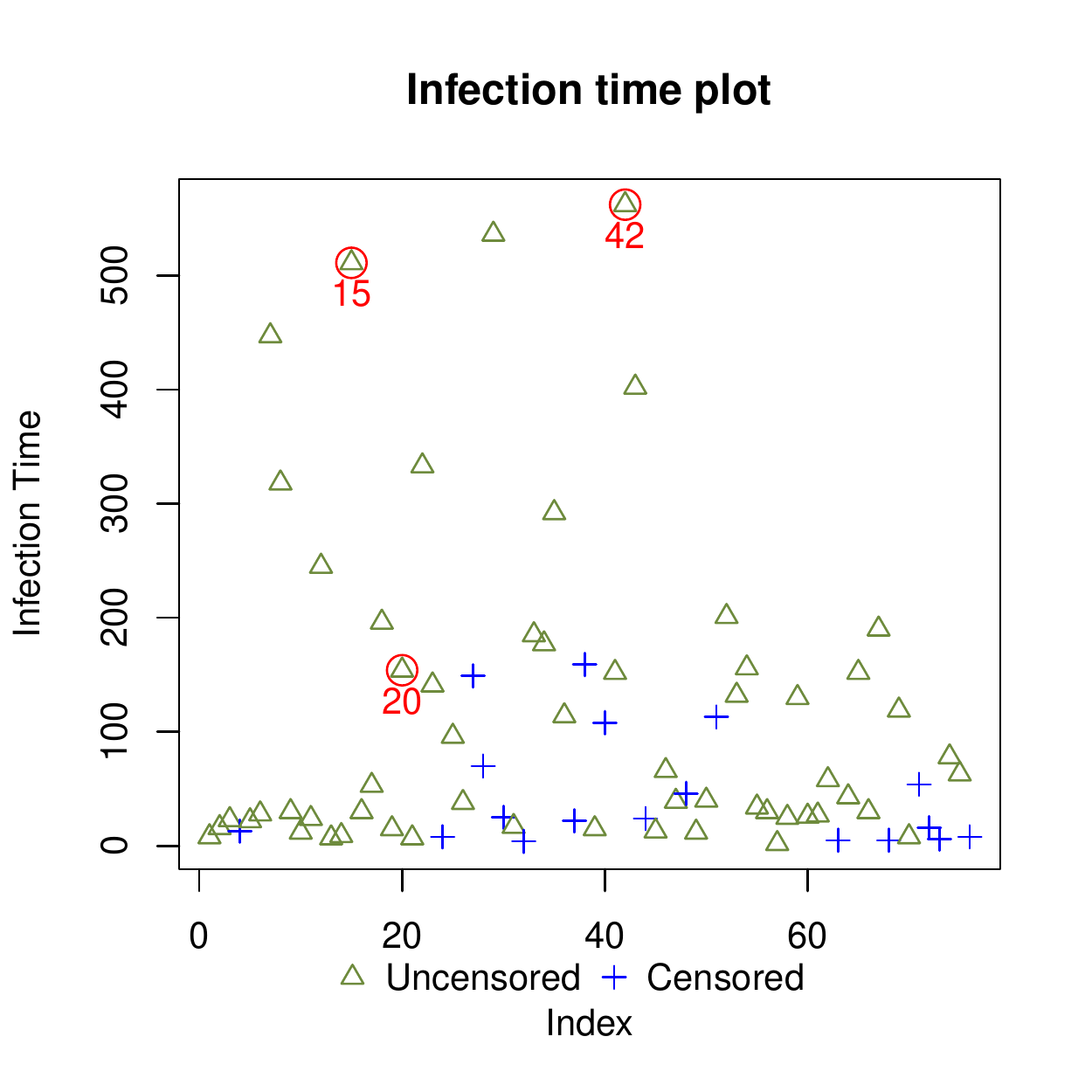}
\caption{The scattered plot of infection times for the kidney infection dataset. }
\label{fig:kidney_timeplot}
\end{figure}

\begin{figure}[htpb]
\includegraphics[width=1\textwidth, height=0.5\textwidth]{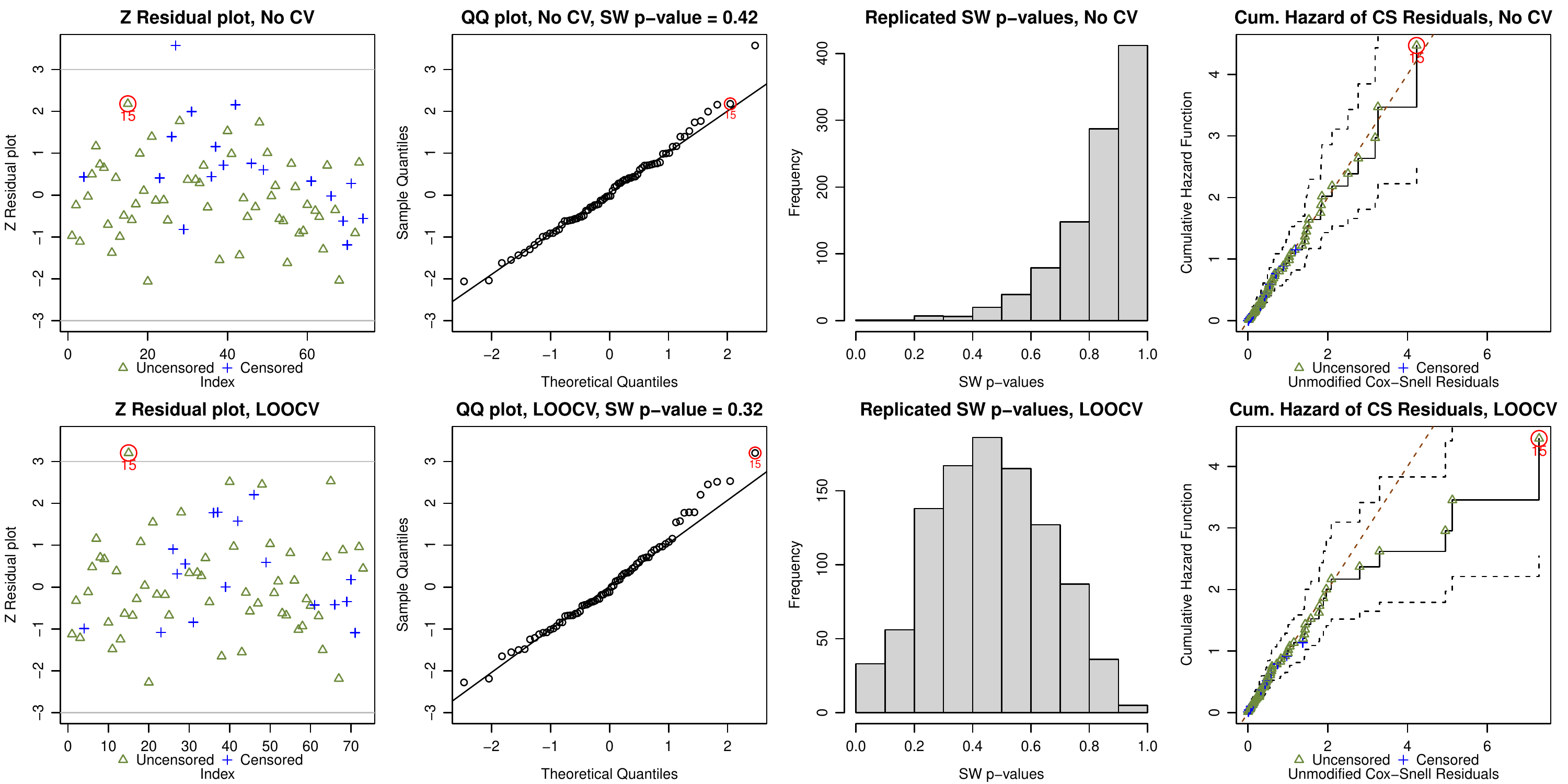}
\caption{Scatterplots and QQ plots of No-CV and LOOCV Z-residuals of the fitted shared frailty models based on the kidney infection dataset with the cases 42 and 20 removed. The third column presents the histograms of 1000 replicated SW p-values of Z-residuals The fourth column shows the CS residuals computed with the No-CV and LOOCV methods. }
\label{fig:kidney2}
\end{figure}

\begin{figure}[htpb]
\includegraphics[width=1\textwidth, height=0.5\textwidth]{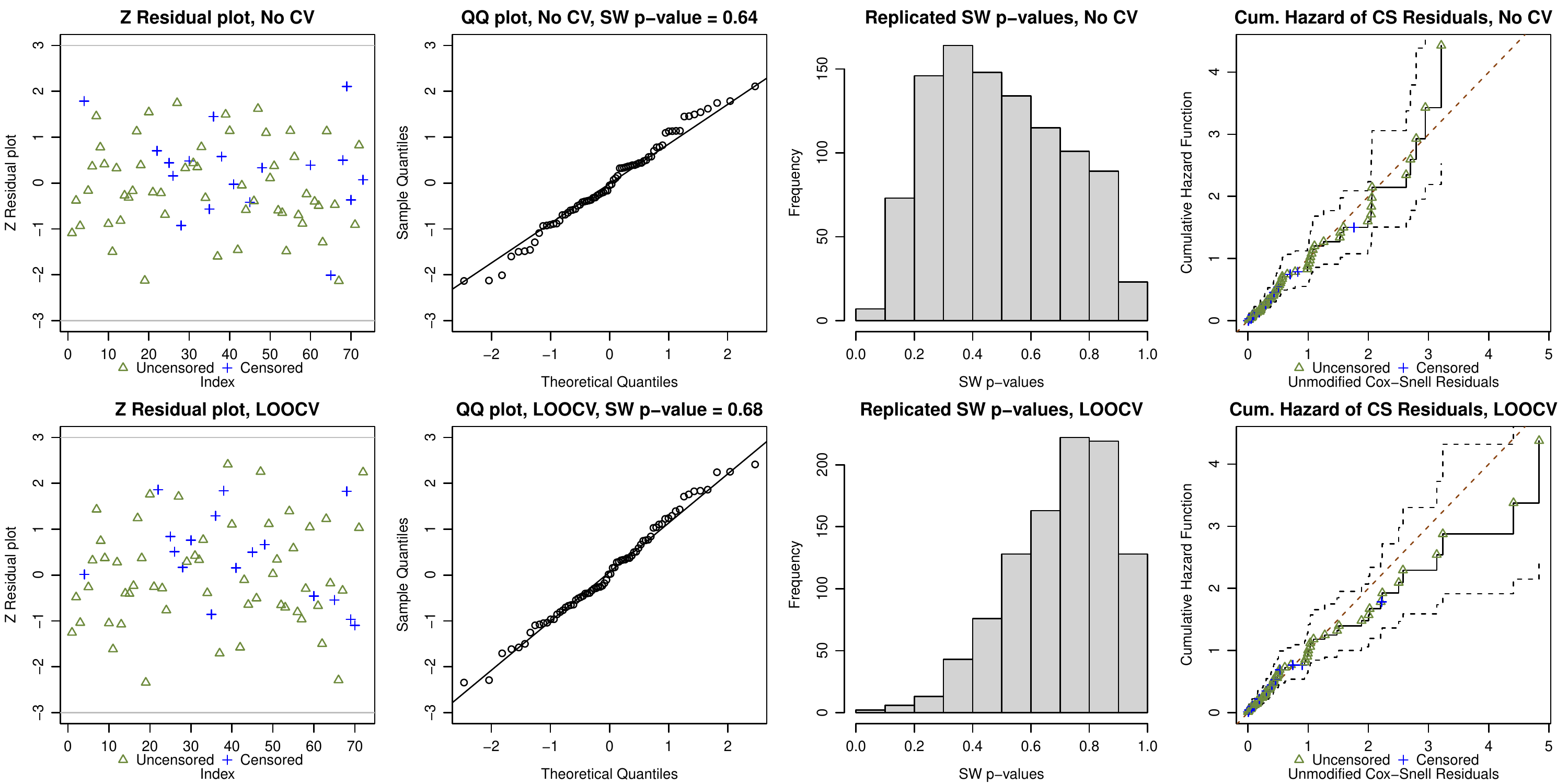}
\caption{Scatterplots and QQ plots of No-CV and LOOCV Z-residuals of the fitted shared frailty models based on the kidney infection dataset with the cases 42, 20, and 15 removed. The third column presents the histograms of 1000 replicated SW p-values of Z-residuals The fourth column shows the CS residuals computed with the No-CV and LOOCV methods.   }
\label{fig:kidney3}
\end{figure}

\newpage

\section*{Acknowledgements}

The authors would like to acknowledge the support from the individual discovery grants awarded to Cindy Feng and Longhai Li by the  Natural Sciences and Engineering Research Council of Canada (NSERC).

 \bibliographystyle{elsarticle-num} 
 \bibliography{cas-refs}





\end{document}